%% file: sandscape_main.tex
\documentclass[%
 reprint,
 amsmath,amssymb,
 aps,
]{revtex4-2}

\usepackage{graphicx}
\usepackage{float}
\usepackage{dcolumn}
\usepackage{bm}

\usepackage{standalone} 

\newcommand{\norm}[1]{\left\lVert #1 \right\rVert}

\usepackage{capt-of} 
\usepackage{graphicx}
\usepackage{multirow}
\usepackage{hyperref}

\begin{document}

\let\oldaddcontentsline\addcontentsline
\renewcommand{\addcontentsline}[3]{}

\preprint{APS/123-QED}

\title{ Sandscapes: self-modifying energy landscapes with emergent branching and flips}

\author{Nacer Eddine Boukacem}
\affiliation{D\'epartement de Biochimie et Medecine Mol\'eculaire, Universit\'e de Montr\'eal, Montr\'eal, QC, Canada\\
Institut Courtois d'Innovation Biom\'edicale, Montr\'eal, QC, Canada
}%

\author{Madhav Mani}
\affiliation{Engineering Sciences and Applied Mathematics, Northwestern University, Evanston, IL, USA\\
NSF-Simons Center for Quantitative Biology, Northwestern University, Evanston, IL, USA
}
\author{Paul Fran\c cois}
 \email{paul.francois@umontreal.ca}
\affiliation{%
D\'epartement de Biochimie et Medecine Mol\'eculaire, Universit\'e de Montr\'eal\, Montr\'eal, QC, Canada\\Institut Courtois d'Innovation Biom\'edicale, Montr\'eal, QC, Canada\\
 MILA Qu\'ebec, Montr\'eal, QC, Canada 
}%

\date{\today}

\begin{abstract}
Energy landscapes provide a common framework for describing learning, embryonic development, and collective dynamics. Although such landscapes may evolve over time, their dynamics are typically prescribed externally rather than generated by the system itself. Here we get inspiration from biology to introduce sandscapes : self-modifying landscapes in which the motions of interacting agents continuously reshape the landscape that governs their own trajectories. We derive sandscapes from a minimal model of interacting Hopfield units, where the basins of each attractor are modulated by their occupancies. Sandscapes spontaneously generate sequential symmetry breaking and differentiation trees, with local branching described by coupled Ising dynamics. We then drive the dynamics of sandscapes (using local proliferation common in biology) and leverage catastrophe theory to show that sandscapes self-organize toward flip bifurcations, suggesting a generic mechanism for the emergence of ubiquitous binary cell-fate decisions. We further demonstrate that sandscapes can act as generative models of developmental trajectories : starting from terminal states alone, we reconstruct realistic hematopoietic differentiation trees with multiple layers of intermediate progenitor states. More broadly, our results identify sandscapes as a general principle of adaptive dynamics, explaining how feedback between agents and landscapes produces branching, criticality, and self-organization across learning and biology.
\end{abstract}

\maketitle

Complex systems such as neural networks, developing tissues or differentiating cell populations must reliably generate structured sequences of fate decisions from high-dimensional, noisy, and interacting states. To understand such dynamics in an embryonic development context, Waddington introduced the concept of epigenetic landscape  \cite{Waddington1957}, wherein different cellular states correspond to different valleys, sequentially splitting to give rise to branching differentiation trees and transitions between cell identities. Recent experimental works on cellular differentiation have essentially validated Waddington's intuition \cite{Saez2021, fontaine2025dynamic,Stuart2026.06.06.730555}, motivating multiple works to map and model the epigenetic landscape. Examples include pioneering work using classical Hopfield networks \cite{Lang2014}, then geometric models \cite{corson2012geometry,corson2017gene,camacho2021quantifying, raju2024geometrical} , via the introduction of epigenetic variables self-regulating to induce bifurcations \cite{Matsushita202, plugers2026noise},  and more recently through explicit landscape models built from data using various mathematical, computational or machine learning methods \cite{karin2024enhancernet, howe2025learning, cislo2025reconstructing, yampolskaya2025hopfield, mochulska2025generative,fontaine2025dynamic, duddu2025distinct}. Importantly, careful mathematical and experimental study of the sequence of cellular decisions have led to the identification of new generic features of (biological) decision, such as the binary (or heteroclinic) flip \cite{Rand2021}, which has now been repeatedly identified in multiple contexts \cite{Saez2021, raju2024geometrical, fontaine2025dynamic, Stuart2026.06.06.730555}. However, despite extensive molecular characterization, the dynamical principles underlying the emergence of specific landscapes remain unclear. In particular, it is not known how collective interactions can generate both stable attractor-like fate identities and reproducible sequences of transitions between them without well adjusted regulatory control.  While such approaches capture aspects of stability, they do not explain how the structure of this landscape itself may emerge from, and evolve with an interacting population. 

Inspiration is offered by machine learning. Associative memory models in statistical physics and machine learning, most notably Hopfield networks and their modern extensions \cite{Hopfield1982,Krotov2016}, provide a mathematical framework for understanding how high-dimensional systems can store and retrieve discrete patterns as attractors of an energy landscape. Generalized Hopfield models (GHM) are a particular class of energy based models, connected to transformers \cite{Ramsauer2020}, and in their simplest form equivalent to diffusion models now used for image generation \cite{krotov2023new, pham2025memorization}. They are versatile enough to lead to concrete applications \cite{Ramsauer2020, widrich2020modern, yampolskaya2025hopfield}, while amenable for more theoretical study.  Recently, Krotov and Hopfield introduced an architecture where Hopfield units are used as the first layer of a classifier model \cite{Krotov2016,Krotov2017}, which, among other properties, offers a direct and intuitive glimpse on how information is encoded and changes during learning \cite{boukacem2024waddington}.  More precisely,  Hopfield units sequentially localize around saddles of trajectories (looking like mixtures of samples in the MNIST context), then sequential symmetry breakings occur so that initially identical units follow gradually specializing paths, thus very reminiscent of the Waddington caricature \cite{boukacem2024waddington}.  This suggests a connection between learning dynamics in Hopfield networks and cellular differentiation \cite{boukacem2024waddington, karin2024, Karin2026}, motivating the present study.

We posit that, in all those examples, interacting agents (neurons, cells) evolve in state space while simultaneously reshaping their own effective dynamical environment.  Inspired by a broad class of biological examples  (Fig. \ref{fig:fig1}A), we use a model of interacting Hopfield units to introduce the concept of sandscape: a self-modifying energy landscape in which the trajectories of interacting agents and the landscape they explore continuously co-evolve. In contrast to conventional time-dependent landscapes (sometimes called seascapes \cite{mustonen2009fitness,blumenthal2026}), whose evolution is prescribed externally, sandscapes emerge from feedback between the evolving distribution of agents and the effective geometry of the landscape they collectively define. Just as grains of sand collectively sculpt dunes while simultaneously being guided by their shape, interacting units reshape the energy landscape as they move through state space, thereby altering one another's future trajectories.  We show that the feedback we introduce naturally generates sequential symmetry breaking and differentiation trees, with local branching dynamics described by coupled Ising models.  Going further, we leverage catastrophe theory for a three-attractor case to draw the motions of the system in catastrophe space.  Inspired by developmental biology, we then introduce local proliferation in state space, and show how the resulting dynamics of the sandscape self-organize in the vicinity of a heteroclinic flip bifurcation, thus similar to what is typically observed in the dynamics of cellular differentiation. Finally, we describe practical applications to biology as a generative model for tree-like dynamical processes : using as only inputs the final differentiation states of hematopoietic cells, our simple model can recover realistic differentiation trees, including known progenitors, and we discuss those results in the context of current models of cell differentiations. Our work thus unifies different perspectives and fields, providing explanations of why non-trivial dynamics such as critical binary flips might be ubiquitous in complex, self-organized systems, and practically suggesting new landscape-based strategies for (biological) data analysis and modelling.

\section{Model and Methods}

To formulate our model, we gather inspiration across systems (see below, Fig.~\ref{fig:fig1} A), wherein multiple units (denoted $\boldsymbol{x_n}$) individually specialize in a coordinated way, modulated by multiple interactions: 
\begin{itemize}
\item each unit is attracted to stored states (or phenotypes, denoted $\boldsymbol{\xi}_\mu$), akin to the behavior in standard Hopfield models
\item in addition, units mutually interact through shared, mean-field, repressive terms (subsequently called $z_\mu$, defined below) so as to diversify their fates 
\end{itemize}
More precisely, units feel two competing forces: an attraction towards phenotypes 
and a global inhibition where units already committed to a given fate globally suppress 
that same fate in others Fig.~\ref{fig:fig1}A. So the overall landscape seen by individual units is modulated by the positions of every other unit, which, as we will study below, results into a dynamical landscape structure with many non-trivial emergent phenomena (branching dynamics, self-organized flips...).

As said above, our inspiration primarily comes from biology. 
In ant colonies, different castes proportions fluctuate in time, and when one caste proportion is too high, the corresponding pheromone concentration leads to a decrease in the production of new individuals from this same caste (through modulation of the larval feeding) \cite{rajakumar2018social,lillico2017regulation, abouheif2021ant, li2024juvenile, matte2025innovation}. 
For cell differentiation, it is well known that if a given cell population is depleted, e.g. in hematopoiesis, the system collectively reorients differentiation towards the missing fate \cite{schoedel2016bulk,baumgartner2018erk}. In embryonic development, recent works have identified clear feedback between states for proportion regulation \cite{Stuart2026.06.06.730555}. In Generalized Hopfield Networks used for classification, specialization of prototypes is directed towards underrepresented samples through an effective interaction in the loss function \cite{boukacem2024waddington}.

Mathematically, we write
\begin{equation} \label{eq: model}
  \partial_t \boldsymbol{x}_n = \frac{1}{Z_n}\sum_\mu \frac{e^{-\beta \norm{\boldsymbol{\xi_\mu - x_n}}^2 } } 
                                               {\sum_m e^{-\beta \norm{\boldsymbol{\xi_\mu - x_m}}^2 }} \boldsymbol{\xi}_\mu - \boldsymbol{x}_n,
\end{equation}
where $\boldsymbol{x}_n$ (resp.~$\boldsymbol{\xi}_\mu$) is the position of unit $n$ (resp.~phenotype $\mu$) in state-space, and $\beta$ is a fixed parameter controlling the stiffness of the interactions. The numerator $S_{\mu n} = e^{-\beta \norm{\boldsymbol{\xi_\mu - x_n}}^2}$ tends to align $\boldsymbol{x_n}$ towards one of the final phenotypes, while the denominator  

\begin{equation} \label{eq:z}
z_\mu=\sum_m S_{\mu m}=\sum_m e^{-\beta \norm{\boldsymbol{\xi_\mu - x_m}}^2}
\end{equation}

prevents convergence towards already well populated $\boldsymbol{\xi_\mu}$'s. Notice that $z_\mu$'s are mean field variables quantifying collective occupancy of phenotype $\boldsymbol{\xi_\mu}$, and in a biological context would correspond to inhibitory signals (e.g. pheromones in the ant colony case). $Z_n=\sum_\mu \frac{S_{\mu n}}{{z_\mu}}$ is a unit-specific normalization factor. Intuitively, this model is thus analogous to having local attractive Gaussian wells \cite{mochulska2025generative}, but negatively regulated size of each basin as they get more populated. Small additive noise is added to the dynamics of each unit in numerical simulations.

Equivalently, the learning dynamics can be formulated as minimizing the energy :
\begin{equation}\label{eq: energy}
  E(\{\boldsymbol{x}_m\}) = -\sum_\mu \ln \Big( \sum_m e^{-\beta \norm{\boldsymbol{x}_m - \boldsymbol{\xi}_\mu}^2} \Big).
\end{equation}

As written in~\eqref{eq: energy}, the model looks superficially similar to a generalized Hopfield network/diffusion model  \cite{bonnaire2026diffusion, pham2025memorization} (see also \cite{karin2024enhancernet, Karin2026} for an application to enhancer dynamics) with energy
\begin{equation}\label{eq: energy_diffusion}
  E(\boldsymbol{x}) = -\ln \Big( \sum_\mu e^{-\beta(t) \norm{\boldsymbol{x} - \boldsymbol{\xi}_\mu}^2} \Big).
\end{equation}

where units are trained to effectively retrieve stored memories. Comparing our model (Eq. \ref{eq: energy}) with Generalized Hopfield/Diffusion (Eq. \ref{eq: energy_diffusion}), the two  big differences are coming from the nature and location of the $\sum$  terms. The $\sum_m $ term  accounts for multiple units $\boldsymbol{x}_m$, so that multiple Hopfield units are optimized at the same time, while the energy in Eq. \ref{eq: energy_diffusion} only depends on one sample  $\boldsymbol{x}$. The $\sum_\mu$  over the memories is in front of the log in Eq. \ref{eq: energy}  while it is inside of the log in Eq. \ref{eq: energy_diffusion} : in the generalized Hopfield model Eq. \ref{eq: energy_diffusion}, this creates a single landscape, shaped by the stored patterns, in which $\boldsymbol{x}$ evolves, while in Eq. \ref{eq: energy},  this creates interactions between Hopfield units ensuring coverage of the phenotypic space. Indeed, while for $m=1$, there is one single minimum for energy Eq. \ref{eq: energy}, corresponding to the average of all patterns, when $m>1$, $\beta$ high enough, it becomes more advantageous for the system to spread the units onto multiple phenotypes. In particular, in the low-temperature limit $\beta \rightarrow \infty$, it is easy to see that $E(\{\boldsymbol{x}_m\}) \sim \beta \sum_{\mu}  min_m \norm{\boldsymbol{x}_m - \boldsymbol{\xi}_\mu}^2$, clearly indicating that the system will tend to localize at least one Hopfield unit at each phenotype. Finite $\beta$ dependency adds 'soft' constraints on the system. All of this greatly influences the dynamics of the system, as we will show below.

Concretely, running the ODE dynamics, as Hopfield units flow through the energy landscape, they distort it (via signal fields $z_\mu$), 
in turn affecting their paths and ultimately their fate. Altogether, each trajectory shapes, and is shaped by, the collective 
motion. For finite $\beta$, typical dynamics proceeds through sequential differentiation, where each split lies along a 
low-dimensional subspace, reproducing the observation of \cite{boukacem2024waddington}. In the three-phenotype case (Fig.~\ref{fig:fig1}B),
units initially converge to a central saddle before undergoing two successive 
binary splits: pink-fated units first separate from the rest, while green and blue-fated 
units move together towards a second saddle where they finally differentiate along an orthogonal direction.

Figure~\ref{fig:fig1}C, illustrates the dynamical nature of the energy landscape:
in early timepoints (i - vi), as units move toward the geometric mean, they are chased by a saddle point which eventually catches up to the undifferentiated group, 
initiating the first differentiation event. In the process, the landscape changes drastically : going from a double well to a triple well potential 
as units differentiate (see Supplementary Movie \ref{mov:Movie 1}, Supplemental Figure~\ref{fig:SFig-dynamical-flow}). Figure~\ref{fig:fig1}D illustrates a 2D representation of what happens when running this formalism on the MNIST dataset, where each datapoint is considered as a final state $\xi_\mu$ : strikingly, we see branching dynamics similar to what we observed before using Hopfield based classification networks \cite{boukacem2024waddington},  with qualitatively identical subtrees (ie. subtrees grouping 4-7-9 and 3-5-6-8), Figure~\ref{fig:fig1}E and Supplementary Movie~\ref{mov:Movie 2}.

\begin{figure*}[htbp]
    \centering
    \includegraphics[width=0.99\textwidth]{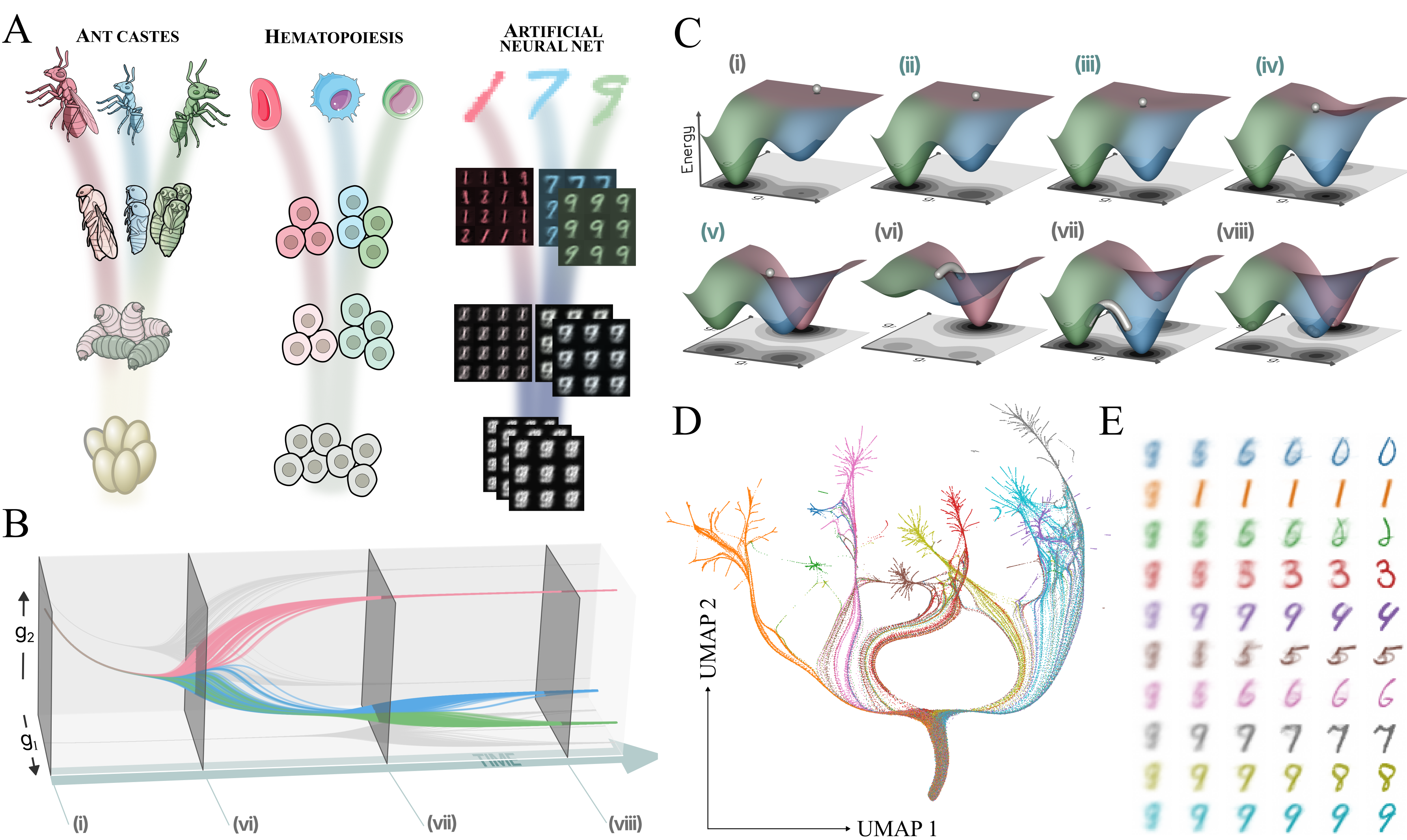}
    \caption{(A) Illustration of the model with three available phenotypes. 
    Each unit continuously emits three phenotype-specific signals, as a function of its position in state space. These signals globally inhibit units from differentiating into the matching phenotype, effectively shrinking the corresponding basin of attraction.
    (B) Differentiation of 500 units across three phenotypes ($\xi_{green}=(-1, -1.3)$, $\xi_{blue}=(1.0, -1.3)$, and $\xi_{red}=(0, 1.3)$). 
    Lines correspond to cell trajectories, with time represented in the $x$-axis, while the vertical plane represent the state space ($g_1$, $g_2$).
    Each line is colored as function of position in state space, illustrating the relative signal emitted by the corresponding unit. 
    (C) Snapshots of the energy landscape and unit positions, (i, vi, vii, viii) corresponding to the gray panels in B, 
    while (ii)-(v) provide additional snapshots of the undifferentiated dynamics.  
    As the simulation progresses, the initial double well potential undergoes a sequence of bifurcations, giving rise to a third attractor, saddles and a source. The dynamics of the energy 
    landscape are driven by effective signals emitted by units, which in turn drives unit dynamics. 
    (D) Snapshot of the dynamics with $\boldsymbol{\xi}_\mu$-phenotypes defined by MNIST digits, and $\boldsymbol{x}_n$ units (400) initialized near the mean digit. Individual points represent $\boldsymbol{x}_n(t)$ vectors projected onto a 2D UMAP embedding for visualization and colored by their terminal fates. (E) Snapshots of ten $\boldsymbol{x}_n$ units, unflattened into 28x28 images and stacked into rows. Each column illustrates the same set of units at different times of the dynamics. }
    \label{fig:fig1}
\end{figure*}

\section{Results}
\subsection{Branching dynamics for three phenotype case}

\begin{figure*}
    \centering
    \includegraphics[width=0.8\textwidth]{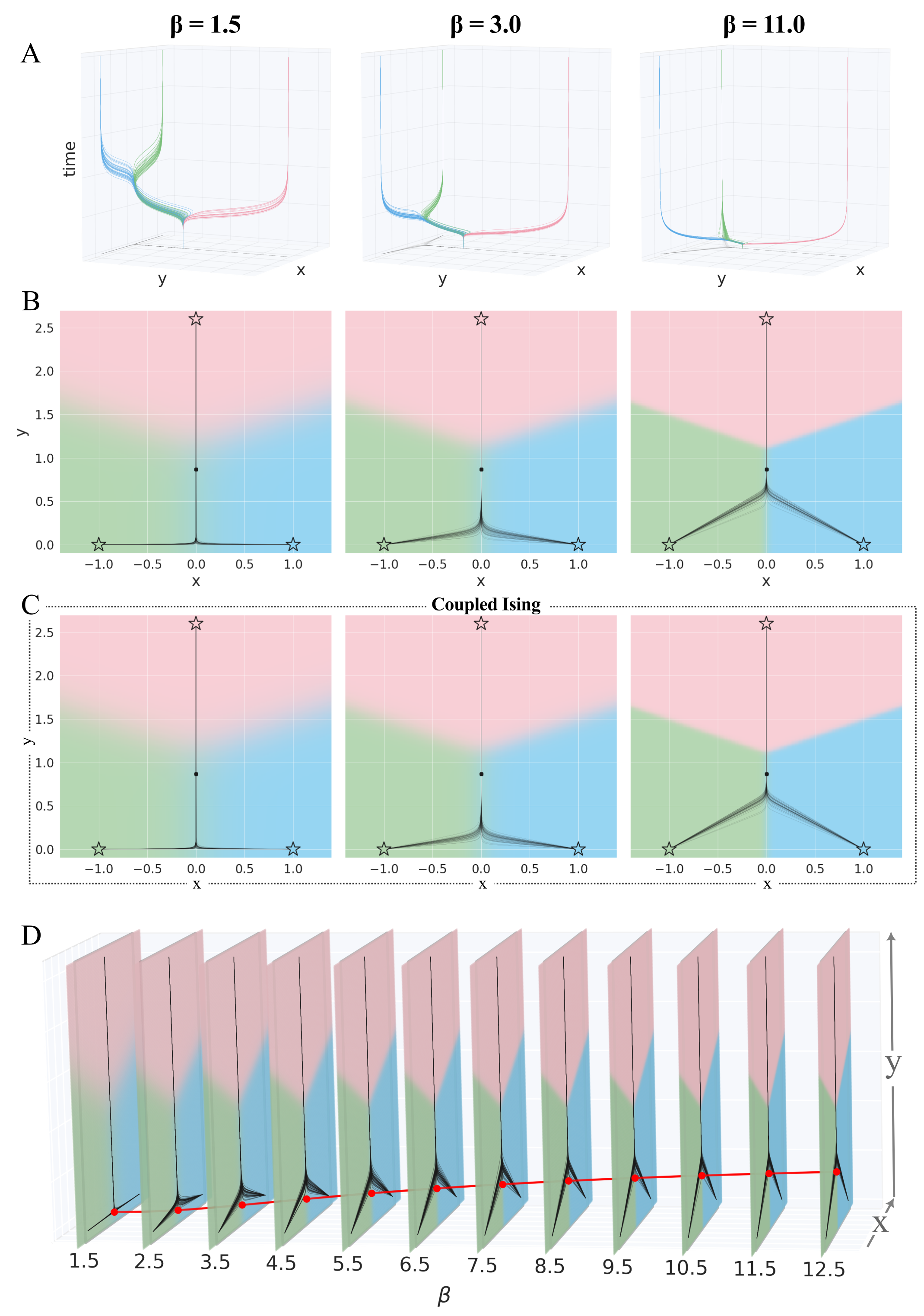}
    \caption{Differentiation dynamics in a three-phenotype system for various $\beta$ values with phenotype geometry $d_x = 2.0$ and $d_y=2.6$. (A) Dynamical trajectories for various $\beta$. The $x$-$y$ plane represents the state space, while the $z$-axis represents time. (B) Dynamical traces projected into the 2D state space, corresponding to (A). At lower $\beta$, differentiation along the $y$-axis happens significantly faster than along the $x$-axis, resulting in a hierarchical sequence of 1D splits. At higher $\beta$, timescales are not well separated, leading to simultaneous differentiation towards all phenotypes. (C) A coupled Ising reformulation of the model recapitulates the results of (B). (D) Dynamical traces for a wider range of $\beta$, illustrating the separation of timescales via the location of the intermediate progenitor state. In red, an analytical approximation of the position of the intermediate progenitor as function of $\beta$.} 
    \label{fig:fig2}
\end{figure*}

Our previous works on coupled Hopfield units \cite{boukacem2024waddington} suggested that, even in a very high-dimensional space, the learning dynamics are Waddington-landscape like with sequential branchings of internal units along distinct 1-dimensional subspaces. The cumbersome nature of the initial models (combining Hopfield units with classifications and very steep non-linearities) did not allow us to fully capture and understand analytically what is happening, and we now revisit the branching problem using the current model.
We study the minimal case exhibiting two sequential branching dynamics: three phenotypes arranged in a tall isosceles \mbox{triangle} (see Fig.~\ref{fig:fig2} A-B, star-shaped markers illustrate phenotype position), with an arbitrary number of units.
Three-phenotype systems always define a stable 2D subspace containing all critical points, without loss of generality, we work in this plane and take as phenotypes :
\begin{gather}
  \boldsymbol{\xi}_1 = (-\frac{d_x}{2}, 0), \quad 
  \boldsymbol{\xi}_2 = (\frac{d_x}{2}, 0), \quad
  \boldsymbol{\xi}_3 = (0, d_y) 
  \end{gather}
where $d_x$ and $d_y$ fully determine our phenotype geometry, and $d_y > \frac{\sqrt{3}}{2} d_x$ injects asymmetry allowing for the generic situation of sequential differentiation (vs a three way branching point). 
In this stable plane, we can show (see Supplement \S\ref{section: Supp Ising}) that the dynamics of every Hopfield unit $\boldsymbol{x_n} = (x_n, y_n)$ reduces to two  coupled Ising mean-field processes : 
\begin{subequations}\label{eq: ising}
  \begin{align}
    \partial_t x_n &= \tanh(\beta d_x x_n - \frac{1}{2}\ln(z_2/z_1)) \frac{k(x_n,y_n) d_x}{2} - x_n, \\
    \partial_t \tilde{y}_n &= \tanh(\beta d_y \tilde{y}_n -\frac{1}{2}\ln(z_3/z_1) - \phi(x_n,z_1/z_2)) \frac{d_y}{2} - \tilde{y}_n,
  \end{align}
\end{subequations}
setting shorthand \mbox{$\tilde{y}_n = y_n - \frac{1}{2}d_y$}, (see Fig.~\ref{fig:fig2}).  Functions $k,\phi$ are defined in Supplement \S\ref{section: Supp Ising}. Notice there are $\ln(z_\mu)$ terms playing the role of biasing fields, that depend on the position of all other Hopfield units. In particular they cancel out if $z_\mu$s are equal, meaning that those biases play a role only when there is an imbalance in the coverage of phenotypes.
Panel B shows trajectories of the original system, while panel C illustrates trajectories for the Ising-like formulation, showing perfect agreement between the two models.

To understand intuitively what happens, consider the dynamics of units initialized near the barycenter of the triangle in the limit of an uncoupled, static, and unbiased landscape (which corresponds to $k(0,0)=1$, $z_1=z_2 \approx z_3$, $\phi \approx 0$). In such case, units feel no \textit{relative} inhibition from their surroundings.

Just like classical Ising mean-field systems, there is a critical $\beta_{C, i} = 2/d^2_i$ below which the barycenter  is stable and units do not differentiate along the $i$-th dimension. For $\beta$ exceeding this critical value, the central fixed point is destabilized, and unit trajectories are sigmoidal : they grow exponentially near the origin, where $\tanh$ is quasi-linear, then follow exponential relaxation as $\tanh$ saturates. This is visible in Fig.~\ref{fig:fig2}A as an S-shaped curves along the time axis (see Supplemental \S\ref{section: Supp piecewise Ising} and Supplemental Fig.~\ref{fig:SFig1}).

In the initial explosion, the exponential growth rate diverges near the critical threshold as $\tau_i \approx {(\beta/\beta_{C, i} - 1)}^{-1}$. As a result, small differences in $\beta_{C, x}$ and $\beta_{C, y}$, which are set by the geometry of phenotypes,  have an amplified effect on the separation of $\tau_x$ and $\tau_y$ near criticality. Here, since $d_y>d_x$, the system breaks symmetry in the $y$ direction first, giving rise to a first split along the $y$ direction. The second split along the $x$ direction occurs after $\tau_x$, and the location of the split in the $(x,y)$ plane depends on the $y$ evolution during $\tau_y$ and $\tau_x$. This location can be calculated using piece-wise linear approximations (see Supplement \S\ref{section: Supp Timescale}), and Fig.~\ref{fig:fig2} D illustrates the  calculated location of the split as a function of $\beta$ along with actual trajectories (see also Fig.~\ref{fig:SFig-progenitor-1}, \ref{fig:SFig-progenitor-2} for a comparison between the analytical approximation and the measured location of the split, showing near perfect agreement for $d_y \gg d_x$).

In particular, we see that in the low $\beta$ limit, the second split happens right between the two phenotypes on the $x$ axis, giving rise to sequential 1-D differentiation, similar to the heteroclinic flip bifurcation, while high $\beta$ leads to a near-trifurcation event close to the central fixed point.

For later time points, the dynamics are too non-linear for analytical calculations, but clearly for units getting close to phenotypes $\boldsymbol{\xi}_3$,  $ \tilde{y}_n \sim d_y/2$ and the prefactor $k(x_n,y_n)$ goes to $0$ so that for those units, $x_n\rightarrow 0$ ensuring that they stabilize close to $x_n=0, y_n \sim d_y$, i.e. $\boldsymbol{\xi}_3$ . Conversely, for units getting close to $ \tilde{y}_n \sim -d_y/2$, the other fixed point for $y_n$, there is some bistability for $x_n = \pm \frac{d_x}{2}$ ensuring eventual convergence close to phenotypes  $\boldsymbol{\xi}_{1,2}$. As said above, the term $\ln(z_2/z_1)$ in the $x_n$ equation ensures that if one of the phenotype is less populated, the overall dynamics is tilted towards it.

\subsection{Differentiation and self-regulation as trajectories in catastrophe space}

In the previous section, we focused on the branching dynamics for a case where the relative coverage of phenotypes (encoded by $z_\mu$s) is essentially unchanged. Yet, on longer time scales, the coverage/energy~\eqref{eq: energy} does not provide a fixed landscape, but what we describe as a \textit{sandscape} dynamically shifting with the motion of every unit. We now consider what happens when the  $z_\mu$s are changing, to capture the full extent of the dynamical fluctuations offered by this sandscape.

Thanks to the introduction of the unit-specific normalization factor $Z_n$, the state of the sandscape can be described with fewer variables than phenotypes using \textit{relative} inhibition.

Specifically, for the three-phenotype case the two ratios $z_1/z_3$ and $z_2/z_3$ ( which for conciseness we write as $\tilde{z}_1$ and $\tilde{z}_2$ ) fully describe the signaling field felt by individual units in 2D and consequently, the instantaneous dynamical geometry of the system.
Fundamentally, $\tilde{z}_1$ and $\tilde{z}_2$ represent relative inhibition of their corresponding phenotype. As such, increasing one while holding the other constant will, in the limit, annihilate the corresponding attractor. Likewise, decreasing both has the reciprocal effect, and suppresses the attractor corresponding to $\xi_3$.  In other words, $\tilde{z}_\mu$ act as control parameters of the sandscape. This allows us to draw catastrophe diagrams (see Fig.~\ref{fig:fig3_1}A, B), where each point in $\tilde{z}$-space is distinguished by the number of available fixed points.

We first illustrate what happens with the tall isosceles case similar to the previous section. We draw the number of fixed points as a function of a  the $\tilde{z}_\mu$s, deriving the catastrophe space. Remarkably, we recover
 the elliptic umbilic catastrophe \cite{Rand2021}, marked by four distinct regions, associated with 7, 5, 3, and 1 fixed point(s) respectively. 
As expected from classical catastrophe theory, the 7-FP region ( a warped triangle centered near $\tilde{z}_\mu \approx (100, 100)$ ) encloses the dynamical topology shown in Fig.~\ref{fig:fig3_1}B (7): characterized by a central repeller, three attractors and three saddles (located between attractors). Encapsulating it, is the teardrop-shaped 5-FP region (Fig.~\ref{fig:SFig2}) whose dynamical topology (see Fig.~\ref{fig:fig3_1}B (1-6)) is characterized by three attractors and two saddles, whose connections are underpinned by heteroclinic flip bifurcations (Fig.~\ref{fig:fig3_1}B (1), (3), (5)). The remaining regions are derived from the annihilation of attractor(s) through saddle node bifurcations; the 3-FP region contains an effectively 1D topology, made of two attractors and one saddle (see Fig.~\ref{fig:fig3_1}B (8, 9) and Fig.~\ref{fig:SFig2}), while the minimal 1-FP region encloses single attractor topologies, illustrated in Supplemental Fig.~\ref{fig:SFig2}.

In our framework, the motion of Hopfield units ($\boldsymbol{x}_n$) in state space in turn changes the values of the $\tilde{z}_\mu$s, giving rise to a self-organized motion in the catastrophe space discussed above.
Revisiting the simulations presented in Fig.~\ref{fig:fig1}C in this lens, initial conditions lead to low relative signals $\tilde{z}_1$ and $\tilde{z}_2$, putting the system in the lower-left corner (3-FP) of catastrophe space, Fig.~\ref{fig:fig3_1}D. 

Early state space dynamics are then accompanied by a nearly linear trajectory in catastrophe space, as $\tilde{z}_1$ and $\tilde{z}_2$ quickly increase together, Fig.~\ref{fig:fig3_1}D (ii-vi). The system first reaches the 5-FP region, with a saddle-node bifurcation as the inter-region boundary is crossed, Fig.~\ref{fig:fig3_1}D (iii) and Fig.~\ref{fig:fig1}C (iii), then the 7-FP region following a pitchfork bifurcation, Fig.~\ref{fig:fig3_1}D (v) and Fig.~\ref{fig:fig1}C (v). 
Remarkably, as units differentiate along the $y$- then $x$-axis, the system drifts back and forth within the 7-FP region, Fig.~\ref{fig:fig3_1}D (v-vii). As such, differentiation of Hopfield units arises spontaneously through the interplay between catastrophe and state space motion, without the need for external input.

\subsection{Perturbing the sandscape I : units removal}

The final state of the system when the number of units is fixed corresponds to a homeostatic state where inhibitions between phenotypes equally balance. In biology, differentiating systems are interesting beyond the homeostatic case, as their response and robustness to perturbations are crucial. For instance, an ant colony that cannot replace lost castes collapses, and a hematopoietic system that cannot recover from an infection is fatal. Conversely, some phenotypes might also proliferate in response to external signals, or simply because they correspond to stem progenitors.  From a dynamical systems standpoint, since sandscapes depend on the location of all units, we expect that targeted removal and introduction of new units should drive the sandscape in characteristic regimes out of homeostatis. With this in mind, we now study how our model of interacting Hopfield units responds to perturbations, employing the prism of catastrophe space to characterize the associated dynamical regimes.

First, we simulated a system with initial conditions as in Fig.~\ref{fig:fig1}C, with the addition of perturbation to unit populations during a short time window $t\in[700, 750]$, namely the removal of all units populating two of the phenotypes.

\begin{figure*}[htbp]
    \centering
    \includegraphics[width=0.99\textwidth]{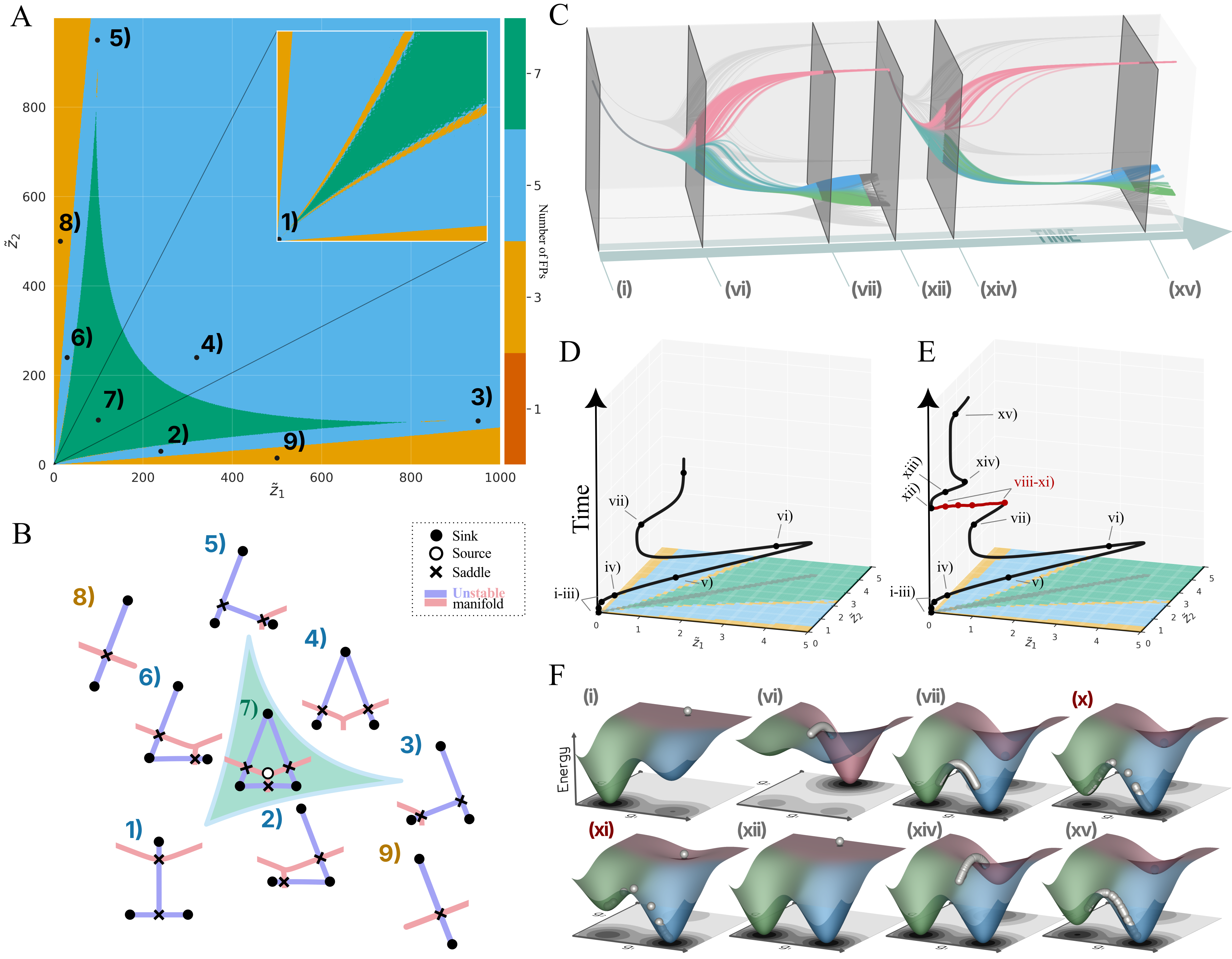}
    \caption{Catastrophe space analysis of the three phenotype geometry, with $\xi_1=(-1, -1.3)$, $\xi_2=(1.0, -1.3)$, and $\xi_3=(0, 1.3)$. (A) The bifurcation/catastrophe space. Colors indicate the number of available fixed points, as a function of the relative signals $\tilde{z}_1$, and $\tilde{z}_2$. (B) Snapshots of dynamical geometries across catastrophe space, corresponding to (1-9) labels in (A). (C) State space dynamics of the differentiating system, perturbed with unit removal. Dynamics are as in Fig.~\ref{fig:fig1}B --- units (500) are initialized near the $\xi_3$ attractor, and left free to differentiate --- except for $t\in [700, 750]$ where units near $\xi_1$ and $\xi_2$ are progressively, and stochastically, pruned until only $\xi_3$-like units remain. (D) Differentiation trajectory of the unperturbed system in catastrophe space, ($\tilde{z}_1(t)$, $\tilde{z}_2(t)$, $t$). (E) Differentiation trajectory of the perturbed system in catastrophe space, where the perturbation (period of unit removal) is highlighted in red. (F) Snapshots of the energy sandscape for points (i-xv) annotated in (D) and (E).}
    \label{fig:fig3_1}
\end{figure*}

The pre-perturbation dynamics are identical to the original system, both in catastrophe-, Fig.~\ref{fig:fig3_1}D, E (i-vii), and state-space, Fig.~\ref{fig:fig3_1}C, Fig.~\ref{fig:fig1}B (i-vii).
Starting at $t=700$, units near the $\xi_1$ (green) and $\xi_2$ (blue) attractors are stochastically pruned, driving the system towards the origin of catastrophe space, Fig.~\ref{fig:fig3_1}E (viii-xi); dying units are marked grey in Fig.~\ref{fig:fig3_1}C. In state space, the energy sandscape, Fig.~\ref{fig:fig3_1}F (x-xi), shows the $\xi_3$ attractor being progressively weakened until ultimately, following successive pitchfork and saddle node bifurcations, it is annihilated, Fig.~\ref{fig:fig3_1}F (xii). 
Post-perturbation, resident $\xi_3$-units find themselves floating over a nonexisting attractor, and begin to re-differentiate to populate the $\xi_1$ and $\xi_2$ phenotypes. This second differentiation process is qualitatively similar to the first: units initially move to the geometric mean (Fig.~\ref{fig:fig3_1}C, F (xii-xiv)), before differentiating sequentially, Fig.~\ref{fig:fig3_1}F (xiv), along the $y$-, and subsequently the $x$-axis, Fig.~\ref{fig:fig3_1}C, F (xv) (see Supplemental Movie~\ref{mov:Movie 3}).

\begin{figure*}[htbp]
    \centering
    \includegraphics[width=0.99\textwidth]{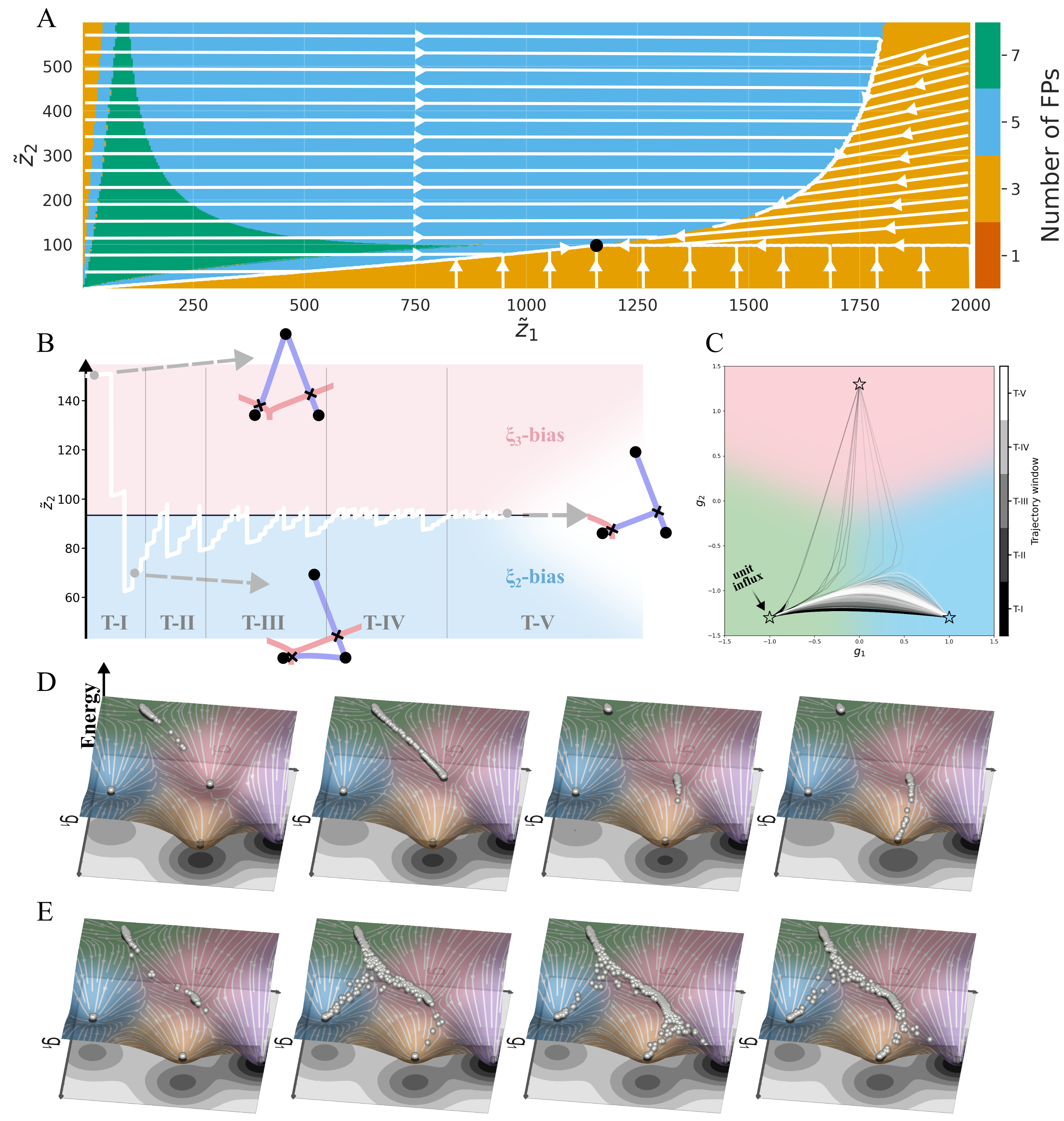}
    \caption{A self-organized heteroclinic flip driven by unit influx for the three phenotype geometry ($\xi_1=(-1, -1.3)$, $\xi_2=(1.0, -1.3)$, and $\xi_3=(0, 1.3)$). (A) Expanded view of the bifurcation / catastrophe space shown in Fig.~\ref{fig:fig3_1}A. White flow lines correspond to the effect the influx of a single $\xi_1$-like unit, with the black point denoting the sole fixed point of the influx-associated flow. (B) Catastrophe trace $\tilde{z}_2(t)$ of a system with constant influx of units (one added unit per 30 epoch) near $\xi_1$, with all units free to move according to \eqref{eq: model}.    
    Snapshots of the dynamical geometry, corresponding to different timepoints of the dynamics, are shown illustrate the dampled oscillation between choice-like geometries and convergence to the critical flip geometry. (C) State space traces of the dynamics, colored by time period (T-I to T-V) as annotated in (B), illustrate the effective canalization of unit trajectories towards the heteroclinic manifold. (D, E) Snapshots of the energy sandscape illustrating a double self-organized heteroclinic flip, for the 5 phenotype geometry ($\xi_1=(-1.75, -1.0)$, $\xi_2=(1.0, -1.5)$, $\xi_3=(0, 1.0)$, $\xi_4=(2.0, 1.0)$, $\xi_5=(0.9, 3.0)$), driven by unit influx near $\xi_1$. (D) Successive snapshots of differentation within the early, choice-like, sandscape. (E) Late snapshots, following spontaneous convergence to the dual-flip geometry.}
    \label{fig:fig3_2}
\end{figure*}

\subsection{Perturbing the sandscape II : Self-organized bifurcation driven by units addition}

Looking at the catastrophe diagram of Fig.~\ref{fig:fig3_1}A, the most prevalent topology in the 5-FP regime is that of the binary choice \cite{Rand2021}, Fig.~\ref{fig:fig3_1}A (2, 4, 6), wherein the two saddles are disconnected from one another. Indeed, the dynamical topology of the system is only ever flip-like, Fig.~\ref{fig:fig3_1}B (i, iii, v), transiently during bifurcations or in very symmetrical cases.

Yet, multiple biological data suggest that flip-like landscapes are very common \cite{Rand2021}, and are often very close to the bifurcation canceling out the 'head attractor', with locally shallow potential \cite{Saez2021, fontaine2025dynamic, Stuart2026.06.06.730555}. While this makes biological sense in a tree-like view of sequential differentiations of cells, this is paradoxical from a dynamical systems standpoint, since this suggests the system is critically tuned close to multiple bifurcations. 

Such biological behaviour can be reproduced and explained within our framework. In biology, progenitors often proliferate, so we consider now a growing system, where an influx of units is added near a single phenotype ($\xi_1$, bottom left of the isoceles triangle). The addition of units changes the value of the $z_\mu$, destabilizing otherwise static regions of catastrophe space, following our sandscape metaphor. The catastrophe space does not generally allow for a dynamical flow  because mean-field signals do not fully determine the system state, however in the case of a slowly growing system, we can obtain useful insight by looking as how the flow field is influenced by the addition of a single unit.

To do this, we consider the motion of a unit in a signaling bath, where $z_\mu$ can be decomposed into a static environmental portion ($z^{\text{env}}_\mu$) and a dynamical term due to an added unit $\boldsymbol{x_{add}}$
\begin{equation}\label{eq: z-decomposition}
    z_\mu = z^{\text{env}}_\mu + z^{\text{add}}_\mu= z^{\text{env}}_\mu + e^{-\beta \norm{\boldsymbol{\xi_\mu - x_{add}}}^2 }.
\end{equation}
The dynamics for this unit can then be written as,
\begin{equation} \label{eq: single}
  \partial_t \boldsymbol{x}_{add} = \frac{1}{Z_n}\sum_\mu \frac{e^{-\beta \norm{\boldsymbol{\xi_\mu - x_{add}}}^2 } } 
                                               {z^{\text{env}}_\mu + e^{-\beta \norm{\boldsymbol{\xi_\mu - x_{add}}}^2 }} \boldsymbol{\xi}_\mu - \boldsymbol{x}_{add},
\end{equation}
with initial conditions close to $\xi_1$, and the environment acting as a large signal reservoir (\mbox{$\sum_\mu z^{\text{env}}_\mu \gg 1$}). As the unit flows within this environmentally biased land/sandscape, it eventually reaches a steady state in state space $\boldsymbol{x}^f_{add}$, and a corresponding one in signaling space, $z^{f}_\mu$.
The latter is then used to compute the change in control parameters following the addition of a single unit, \begin{gather}
        d\tilde{z}_\mu = \tilde{z}^f_\mu - \tilde{z}^{\text{env}}_\mu,
\end{gather}
reflecting the changing land/sandscape due to the addition of one unit. This defines a flow in catastrophe space driven by the addition of one unit (and thus growth). Such flow can be numerically computed by scanning over a grid of environmental conditions, $z^{\text{env}}_\mu$, and is drawn in Fig. \ref{fig:fig3_2} A.

In the 7- and 5-FP regions, the added unit converges to the $\xi_1$ attractor, increasing the relative $\xi_1$ inhibition, generating flow lines pointing to the right in catastrophe space, Fig.~\ref{fig:fig3_2}A. This leads to the eventual destabilization of the $\xi_1$ attractor i.e. towards the 3FP region. Conversely, in the 3-FP region, the absent $\xi_1$ attractor allows an added unit to flow to either of $\xi_2$ and $\xi_3$ depending on the instantaneous balance of signals, hence slowly counteracting the buildup of $\xi_1$ inhibition. Because of this, a system in the 3-FP region is driven back towards the 5 FP region, Fig.~\ref{fig:fig3_2}A, suggesting that it is naturally attracted towards the boundary, thus becoming critical. %

More quantitatively, we indeed see upon full computation of the flow that the system is driven to a critical point along the transition line between the 5FP and 3FP region, where two bifurcations intersect: the annihilation bifurcation of $\xi_1$ and the heteroclinic flip bifurcation between $\xi_2$ and $\xi_3$, black point in Fig.~\ref{fig:fig3_2}A. At this point, units added near the $\xi_1$ attractor are maximally potent: slight imbalances between (1)-(2,3) fates are counteracted by the annihilation or creation of the $\xi_1$ attractor, while imbalances between $\xi_2$ - $\xi_3$ fates are easily recoverable by moving in either direction of the heteroclinic bifurcation. We thus conclude from this analysis that additional growth should naturally drive the system close to this critical point in the catastrophe space.

To check this, we indeed simulated the dynamics of a full system, Eq. \ref{eq: model} where units are slowly added close to $\xi_1$ with some constant rate. Despite the added complexity, this richer setting upholds the insights obtained from the toy model with a single added unit. The system converges to the expected critical point, see Fig.~\ref{fig:fig3_2}B, albeit with overshoots and oscillations depending on the magnitude of noise and the growth timescale. In particular, we see that the saddle close to $\xi_1$ is moving around it (T-III and T-IV periods), biasing the flow towards either $\xi_3$ or  $\xi_2$ in succession. Eventually, the saddle stabilizes and directly connects to the second saddle between  $\xi_3$ and  $\xi_2$  (T-V periods), thus realizing a perfect flip where the released units are bipotent.

As the system grows, added units provide progressively finer control over the fate imbalance, dampening the oscillations between broken choice topologies and eventually locking the system in a flip topology, Fig.~\ref{fig:fig3_2}B. This is illustrated in the time trace of the $\tilde{z}_2$ mean field variable: early dynamics showing $\tilde{z}_2(t)$ alternating around $\approx 95$, with the amplitude of oscillation dying down with time, Fig.~\ref{fig:fig3_2}B. Alternatively, we plot the dynamical traces of units in state space, coloring each trajectory by the time window of birth. Early-born units follow nearly linear trajectories, reflecting the underlying binary choice landscape, see darker trajectories in Fig.~\ref{fig:fig3_2}C. Contrastingly, later-born units first move towards the $\xi_2$ - $\xi_3$ midpoint, before committing to either of the phenotypes, hinting at the saddle geometry of the underlying flip topology, see lighter trajectories in Fig.~\ref{fig:fig3_2}C.

Together these results indicate that local proliferation is enough to drive the system towards a dynamical topology which otherwise would require fine-tuning or inordinate symmetry. Beyond convergence to a heteroclinic flip, we further find that the tug of war between the influx of units and differentiation spontaneously maintains the system at the edge of attractor $\xi_1$ annihilation, maximizing potency of the influx state. This is exactly the kind of behaviours observed for progenitor cells, but as illustrated here, this simply emerges from the very biological assumptions that states  self-repress while one population of progenitors proliferates.

Lastly, we notice that this self-organized state is robust and generic. First, it does not depend on any pre-existing symmetry in the system since we chose the bottom left attractor close to the origin  ($\xi_1$ ) as an influx point, so that $\xi_2$ is much closer to it than $\xi_3$, and accordingly the 'flip' saddle between $\xi_2$ and $\xi_3$ is moving much closer to $\xi_2$ , Fig. \ref{fig:fig3_2}B-C. Second, such self-organized behaviour generalizes naturally beyond three phenotypes, as the growth influx propagates through differentiation, further consistent with the notion of differentiating trees. For instance, in  Fig.  \ref{fig:fig3_2}D-E we show that an asymmetrical arrangement of 5 phenotypes eventually leads to a dynamical topology holding two sequential heteroclinic flips, with both $\xi_1$ and $\xi_3$ at the edge of annihilation Fig.  \ref{fig:fig3_2}E, bottom right. Effectively, differentiation of units from $\xi_1$ to $\xi_3$ acts as an influx, and as such leads to the same topological convergence as described in the three phenotype case, this time between $\xi_3$, $\xi_4$, $\xi_5$, see Fig.~\ref{fig:fig3_2}D, E and Supplemental Movie~\ref{mov:Movie 4}.

\begin{figure*}[htbp]
    \centering
    \includegraphics[width=0.99\textwidth]{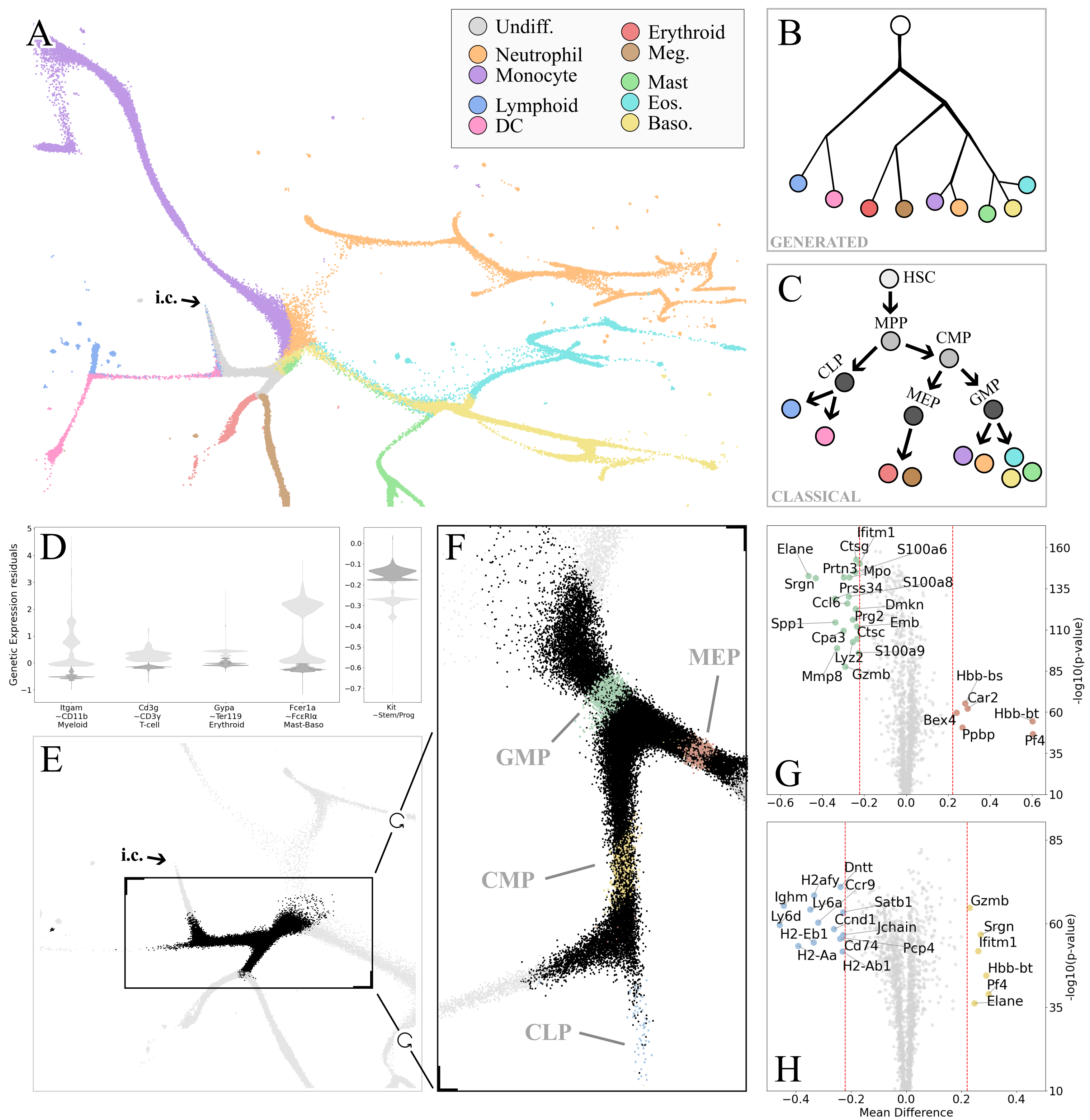}
    \caption{ A data-driven differentiation tree of hematopoiesis. (A) Projection of the differentiation dynamics onto a three dimensional linear embedding defined by Neighborhood Component Analysis (NCA). Points represent individual units, and are colored by their terminal fates, see legend. Unit influx is at a constant rate, 50 added units per epoch, with initial conditions corresponding to the mean stem state, indicated as \textit{i.c.}. (B) Graphical representation of the differentiation topology seen in the dynamics. (C) Graphical representation of the classical mouse hematopoietic differentiation hierarchy, and canonical progenitors. (D) The transcriptomic criteria analogous to $\text{Lin}^-$ (left) and $\text{Kit}^{+}$ (right), used to isolate the central multipotent units highlighted in (E). (E) Snapshot of the dynamics, where multipotent units isolated with (D) are highlighted in black and other units are drawn in grey. (F) Inset of (D) with 4 subclusters isolated by gene expression, analogous to GMP, MEP, CMP and CLP. (G-H) Differential Expression analysis of the progenitor clusters highlighted in (D). Each panel illustrates mean expression difference versus statistical significance for two pairwise comparisons: GMP-MEP (G) and CLP-CMP (H). Negative mean differences, represent genes more highly expresesd in GMP (G) and CLP (H) respectively, while positive differences indicate genes more highly expresesd in MEP (G) and CMP (H). } 
    \label{fig:fig4}
\end{figure*}

\subsection{Data-driven reconstruction of differentiation dynamics}

Thus far, we have dealt with synthetic systems where the phenotype geometry is manually chosen to explore and understand different features of the model. We now integrate biological data into this framework, and show how our formalism can further be used as a generative model for tree-like, differentiating processes. To this end, we draw on single-cell sequencing (scRNA-seq) data, which provides high-dimensional transcriptomic snapshots of individual cells.

We focus on hematopoiesis (the continuous production of blood and immune cells ) as a case study, specifically using a mouse in-vitro dataset \cite{Weinreb2020}. Indeed, blood and immune cells interact through circulating signals, making this differentiation process largely aspatial, hence matching our mean-field signaling perspective. Moreover, hematopoiesis has a well-characterized differentiation hierarchy, providing a ground truth to benchmark the model against. 
In this context, state space corresponds to the 100 principal components (PCs) of  pre-processed gene expression. We adopt a fully unsupervised approach, and define each $\xi_\mu$ as the PCs of exactly one profiled cell.
In order to avoid biasing the model towards more populated cell types, the dataset is subsampled such that cell types are equally represented in the $\{\xi_\mu\}$ set.
Importantly, only terminal cell type information is held in $\xi_\mu$ patterns, progenitor states are not included in the model, with the underlying goal of recovering them within the dynamical structure. In addition, for computational efficiency early Hematopoietic Stem Cell (HSC) samples are not included as patterns, but are averaged over and used as initial conditions. We further assume there is an influx of cells in the growing system close to those initial conditions, thus fully mimicking a differentiating system. 

We run and study dynamics defined by Eq. \ref{eq: model}. In order to visualize the resulting dynamics, we project the 100-dimensional state space onto a three-dimensional linear embedding constructed via Neighborhood Component Analysis (NCA) \cite{NCA_original, sklearn}, see Supplemental Fig.~\ref{fig:SFig4} for a projection of the $\xi_\mu$ patterns in this space. In such visualizations, points are colored by their cell type label, or in the case of simulated data, by the the cell type of the nearest data point at the final timepoint. 
This method allows us to view the global geometry of the dynamics while avoiding non-linear artifacts that may distort our trajectories, or induce discontinuities. 
Initial dynamics show units shooting out of the stem state, in nearly straight trajectories towards all phenotypes, akin to the dark lines in Fig.~\ref{fig:fig3_2}C, see Supplemental Movie~\ref{mov:Movie 5}. 
As the system grows, cell trajectories curve towards emergent intermediate states, and the nozzle of the stem state progressively narrows, eventually establishing a tree visible in Fig.~\ref{fig:fig4}A, which shows a snapshot of the ongoing dynamics in which all units are in continuous motion.

Looking at the dynamical flow of Fig.~\ref{fig:fig4}A-B we see the following differentiation hierarchy. First, HSCs move together towards a saddle  before differentiating into two categories, lymphoid-fated and myeloid-fated cells. 

The first group moves together before differentiating into plasmacytoid dendritic cells (pDCs) and lymphoid cells, while the second group follows a more elaborate differentiation trajectory with multiple successive binary decisions.  

The first binary decision divides the group into two: Megakaryocytes/Erythroids (ME) on one side and Granulocytes/Monocytes (GM) on the other. The ME-group subsequently splits into the individual terminal states, while the GM-group splits into Monocyte/Neutrophils and Mast/Baso/Eos. Monocytes and Neutrophil move to a common saddle before differentiating into their respective terminal states. The granulocyte trio undergoes two further binary events, first separating Mast cells from the group, after which Baso and Eos move together to a late progenitor before finally splitting. 

The differentiation hierarchy inferred from the dynamics is illustrated in Fig.~\ref{fig:fig4}B. Remarkably, this is qualitatively very close to the classical differentiation tree, illustrated in Fig.~\ref{fig:fig4}C, based on the immunophenotypic isolation of multipotent populations \cite{Akashi2000, Adolfsson2005}. 
The very first saddle matches the observed Multi Potent Progenitor (MPP) in hematopoiesis \cite{perie2015branching}, and the subsequent split into lymphoid- and myeloid-fated cells is consistent with the Common Myeloid Progenitor (CMP) and Common Lymphoid Progenitor populations, while the subsequent specification of the myeloid-fated cells into granulocyte/monocyte and megakaryocyte/erythroid is consistent with the Granulocytes Monocytes Progenitors (GMP) and Megakaryocytes Erythroids Progenitor (MEP) populations isolated experimentally. There is clonal evidence suggesting a Monocyte/Neutrophil-restricted progenitor \cite{Weinreb2020}, as well as experimental evidence of Baso/Eos/Mast-restricted progenitors downstream of GMP \cite{Arinobu2005, Hamey2020}, consistent with our derived hematopoietic structure.
Crucially, none of these progenitor stages were provided to the model, the hierarchical structure is inferred from the HSC initial condition and the terminal cell states. In addition, we reiterate that the cell type labels are purposefully not provided to the model either, hence the binary splits (e.g. between Megakaryocytes and Erythroid) are not a result of discrete class identities, but rather emerge from the transcriptional geometry of the profiled cells. 

To confirm if this topological correspondence is matched by actual genes, we project our units back onto genetic expression space using the PCA pseudoinverse, with the goal of isolating predicted progenitors within our tree. To do with, we remove all units expressing lineage specific genes, to isolate the units generated by our procedure which are not terminally differentiated. 
In details, we select for low expression of \textit{Itgam}, \textit{Cd3g}, \textit{Gypa} and \textit{Fcer1a} (chosen as transcriptomic analogs to the experimental lineage (\textit{Lin}) cocktail \cite{LinSpangrude1988} ) to discard most myeloid-, erythroid-, and lymphoid-differentiated units. 
The remaining units are then filtered by \textit{Kit} expression ( a known proxy for potency \cite{KitOkada1991} ) discarding residual myeloids, megakaryocytes, and late lymphoids, which all express \textit{Kit} at low levels.

These transcriptomic criteria are illustrated in Fig.~\ref{fig:fig4}D, and allow us to isolate a central multipotent cluster generated by our procedure, highlighted in Fig.~\ref{fig:fig4}E.
Continuing with the immunophenotype analogs, we use \textit{Slamf1} and \textit{Fcgr3} expression  ( proxies of CD150 and CD16 ) to define four subclusters within the multipotent units, namely presumptive CMP, CLP, MEP and GMP, see Fig.~\ref{fig:fig4}F and Supplemental Fig~\ref{fig:SFig5}.  We then extract the differential expression signature of each branch by pairwise comparison of these subclusters at each binary decision (i.e. CMP vs CLP and MEP vs GMP).

Along the first decision, we find that presumptive CMP units upregulate a mixture of genes spanning their prospective fates, namely erythroid-associated \textit{Hbb-bt}  \cite{HbbGrosveld1987position}, megakaryocyte-associated \textit{Pf4} \cite{Pf4Kowalska2004, Pf4Tiedt2006}, and neutrophil-associated \textit{Elane} \cite{ElaneBelaaouaj1998mice, ElaneHe2026elk1}, with the addition of the pan-hematopoietic \textit{Srgn} \cite{SrgnBrink2004, SrgnZernichow2006, SrgnWoulf2008}. 
Conversely, presumptive CLP units differentially express almost exclusively lymphocyte markers, capturing established hallmarks of early lymphoid differentiation such as \textit{Dntt}, \textit{Ly6d} and \textit{Ly6a} \cite{Ly6D_CLPinlay2009ly6d, DnttCLPMansson2010single, Dntt_klein2022dntt},  the latter consistent with the immunophenotype of CLP \cite{CLPkondo1997identification}. 
Going further down the tree, the second decision shows similar behavior, with presumptive MEP expression spanning both characteristic erythroid (\textit{Hbb-bt}, \textit{Hbb-bs}, \textit{Car2}) and megakaryocyte (\textit{Pf4}, \textit{Ppbp}) genes \cite{EryGenesFraser1987, PpbpMcLaren1982, Pf4Kowalska2004, Pf4Tiedt2006, MouseCellAtlas3Wang2022}, albeit with the unexpected expression of \textit{Bex4}, which we also find expressed in megakaryocyte-differentiated units.  
The presumptive GMP subcluster echoes a similar specific behavior, with expression accross neutrophil- (\textit{Elane}, \textit{Mmp8}), monocyte- (\textit{Lyz2}, \textit{Spp1}) and granulocyte-associated (\textit{Cpa3}, \textit{Prss34}, \textit{Prg2}) genes \cite{ElaneBelaaouaj1998mice, ElaneHe2026elk1, Cpa3Lilla2011, MouseCellAtlas3Wang2022}, see Fig.~\ref{fig:fig4}G and H. 
As a whole, we thus find that the transcriptional signature of these presumptive progenitor clusters matches the idea of multilineage priming \cite{Hirschi2017}, as reflected in the co-expression of genes otherwise associated with distinct downstream fates. Of note, there are well-known controversies on the precise process leading to hematopoiesis \cite{Hirschi2017}, in particular it is not clear how controlled yet artificial in vitro differentiation starting from stem cells \cite{Weinreb2020} matches the in vivo process. For instance, in vivo experiments have shown that at the MPP stage, there is already priming towards either CLP or CMP fates \cite{perie2015branching}. Such behaviour is expected if MPPs indeed correspond to a saddle, since cells on either side of the unstable manifold connecting HSCs to MPPs would be already committed to either lymphoid or myeloid lineage (see e.g. late time trajectories  in Fig. \ref{fig:fig3_2} B-C). Such early priming is one of the main characteristics of heteroclinic flip bifurcations \cite{Rand2021}, thus  naturally occurring in our framework.

\section{Discussion}

The concept of landscape-driven dynamics, coming from physics, has proved productive to describe biological systems ranging from differentiating cells (Waddington's caricature leading to modern catastrophe theory based formalisms \cite{Waddington1957}) to species evolving  (Sewall Wright's fitness landscapes inspiring all modern population genetics \cite{Wright1932}).  Energy based models introduced by Hopfield to first account for associative memory eventually led to practical implementations in machine learning and artificial intelligence, e.g. state-of-the art generative models for pictures rely on Diffusion Models equivalent to Generalized Hopfield models \cite{ho2020denoising,pham2025memorization}. In most of these settings, however, the landscape is effectively exogenous and any temporal variation is imposed externally through changing environments, annealing schedules, or prescribed dynamics \cite{mustonen2009fitness,bonnaire2026diffusion}. Here we consider an opposite limit and introduce a multi-agent energy-based system in which the units evolve within a landscape that is continuously reshaped by their collective state occupancy. We refer to this class of systems as sandscapes: landscapes that both shape and are shaped by the dynamics of their inhabitants. In contrast to conventional landscape models, where topology precedes dynamics, the topology of a sandscape emerges from the dynamics itself. The key feature of sandscapes is a feedback loop in which population occupancy actively reshapes the landscape, effectively turning occupancy variables ($z_\mu$s) into dynamical parameters of the landscape itself. This feedback underlies the emergence of branching, hierarchy formation, and near-critical fate decisions.

A striking feature of sandscape dynamics is the emergence of sequential branching. In numerical examples, and analytically in a mean-field Ising setting, we find that branching arises from the competition between two effects: attraction toward collective fate states, and repulsion between coexisting fates. As a result, initially coherent trajectories split into multiple branches, producing a hierarchical structure without any imposed architectural constraints.  This mechanism differs from branching-like structures observed in diffusion models \cite{biroli2024dynamical} or Generalized Hopfield-based enhancer models \cite{karin2024}, where trajectories evolve in a prescribed time-dependent landscape and diversification is driven by externally imposed temperature schedules. In sandscapes, branching is instead generated endogenously through interactions between coexisting units. In that sense, the dynamics are closer in spirit to ecological diversification, where frequency-dependent competition can destabilize a single phenotype and promote evolutionary branching \cite{dieckmann1999origin}.

This endogenous hierarchy is particularly relevant in the hematopoietic context we practically study. There, the internal branches generated by the model can be interpreted as progenitor states, while the branching endpoints correspond to differentiated fates. Importantly, the intermediates are not static attractors; rather, they are transient states in flux, consistent with lineage priming and  the long-standing “tree versus rainbow” debate in differentiation \cite{Hirschi2017,velten2017human}. Our framework provides a dynamical realization in which discrete lineage branching coexists with continuous progression through transient intermediate states.  From this perspective, progenitor states emerge not as stable attractors but as metastable or saddle-like regions generated by the collective dynamics itself.

Recent quantitative descriptions of cellular differentiation increasingly rely on explicit landscape reconstructions and bifurcation-theoretic descriptions of fate choice \cite{Saez2021,camacho2021quantifying,yampolskaya2025hopfield,fontaine2025dynamic,Stuart2026.06.06.730555}. Catastrophe theory provides a natural language for describing binary fate decisions and their modulation by control parameters \cite{Rand2021}.  In our model, catastrophe-theoretic control parameters become functions of the instantaneous distribution over states, allowing bifurcation structure to evolve dynamically rather than being externally tuned.
This offers a possible mechanism for the robust appearance of near-critical differentiation dynamics, close to a binary flip, as observed experimentally in multiple contexts \cite{Saez2021, fontaine2025dynamic, Stuart2026.06.06.730555}. In static landscapes, the coexistence of a shallow progenitor state and a flip bifurcation would require substantial parameter tuning, as it lies near multiple bifurcation boundaries. In sandscapes, however, this configuration emerges spontaneously from the feedback between proliferation and repression. The resulting dynamics resembles self-organization toward bifurcation points and are suggestive of sandpile-like critical behavior \cite{bak1988self},  further justifying our sandscape analogy. Establishing a genuine self-organized critical regime would however require a dedicated analysis of scaling properties.

We thus predict that cellular differentiation through binary-flip bifurcations could be associated with two generic ingredients: proliferation within some stem-cell or progenitor-like states, and effective self-repression of competing fates. Remarkably, the control parameters governing the bifurcations can be expressed as simple ratios of the $z_\mu$s, providing a natural interpretation of fate control in terms of collective state occupancy, and consistent  with recent experimental evidence of feedback between cell fate proportions and cell decisions \cite{Stuart2026.06.06.730555}.

This raises the question of the biological identity of variables $z_\mu$. Within the model, they act as rapidly communicating occupancy signals that couple the dynamics of distinct cellular states and reshape the effective landscape. Classical diffusible signaling molecules such as FGF, EGF, BMP, and related morphogens therefore constitute natural candidate realizations, e.g. in \cite{Stuart2026.06.06.730555} BMP is indeed secreted by floorplate precursors to influence the earlier neural progenitors vs floorplate precursors flip decision.   With proliferation, progenitor-rich configurations often dominate the dynamics because cells only exit progenitor states past the bifurcation (transiently) canceling it (Fig. \ref{fig:fig3_2} B). While such behavior is broadly consistent with transient developmental contexts \cite{Saez2021}, it is generally not observed in homeostatic tissues such as adult hematopoiesis, where progenitor populations remain relatively small. Additional exogenous sources of such signals could then further modify the landscape, analogous to the decomposition between environmental and endogenous contributions introduced in Eq.~\ref{eq: z-decomposition}, e.g. stem-cell niches may be viewed as localized external sources modulating the corresponding $z_\mu$ independently of the instantaneous cellular composition (see also \cite{mochulska2025generative} for an explicit model of bifurcations generated through modulation of Gaussian attractors).

Broadening the scope of the model, introducing explicit spatial transport or temporal dynamics for the $z_\mu$ would considerably enrich the phenomenology. Even in the present framework, stochastic cell addition produces transient overshoots and relaxation dynamics, Fig. \ref{fig:fig3_2} B. The inclusion of delayed feedback, diffusion, or spatial heterogeneity would naturally generate oscillatory, excitable, or potentially chaotic regimes, reminiscent of the complex temporal dynamics observed in hematopoiesis \cite{jia2024chaotic}.

Our work suggests a view of biological differentiation as a generative process in which the generated population continuously modifies the geometry from which future states are produced. In contrast to conventional generative models evolving within fixed latent structures, sandscapes couple generation and geometry through collective feedback. This perspective naturally connects developmental dynamics with recent work on optimal control of evolving distributions and generative systems \cite{reddy2025dynamic}. The repressive interactions introduced here provide one simple control mechanism, but more sophisticated strategies could be envisioned based on alternative optimization criteria \cite{blumenthal2026}. Beyond their conceptual interest, such approaches may help bridge the gap between modern mathematical methods for evolving probability distributions and realistic biological questions, including progenitor-state identification \cite{salazar2026stochasticity}, fate prediction, and the rational design of differentiation protocols. 

In particular, population occupancy becomes a dynamical control parameter for landscape geometry. In this view, differentiation is not the exploration of a pre-existing developmental landscape, but the collective construction of that landscape by the cells themselves.

\section*{Numerical Methods}
All codes used for this paper are publicly available at \url{github.com/nacer-eb/Sandscapes}, with figure-specific executable codes allowing for full figure generation.

\bibliography{sandscape}

\clearpage
\onecolumngrid 

\begin{center}
    \textbf{\large Supplemental Material}\\[.2cm]
\end{center}
\vspace{1cm}

\setcounter{section}{0}
\setcounter{equation}{0}
\setcounter{figure}{0}
\setcounter{table}{0}
\setcounter{page}{1}

\renewcommand{\thesection}{S\arabic{section}}
\renewcommand{\theequation}{S\arabic{equation}}
\renewcommand{\thefigure}{S\arabic{figure}}
\renewcommand{\thetable}{S\arabic{table}}

\makeatletter
\renewcommand{\l@figure}{\@dottedtocline{1}{0em}{2.5em}}
\makeatother

\tableofcontents
\listoffigures
\vspace{1cm}

\let\addcontentsline\oldaddcontentsline

\clearpage

\input{paper_supplement}

\end{document}

%% file: paper_supplement.tex
\section{The Ising formulation} \label{section: Supp Ising}
In this section, we detail the derivation of the Ising formulation from eq.~\eqref{eq: model},
\begin{equation}
    \partial_t \boldsymbol{x}_n = \frac{1}{Z_n} \sum_\mu \frac{e^{-\beta \norm{\boldsymbol{\xi_\mu - x_n}}^2}}{ \sum_m e^{-\beta \norm{\boldsymbol{\xi_\mu - x_m}}^2} } \boldsymbol{\xi}_\mu - \boldsymbol{x}_n,
\end{equation}
with $Z_n = \sum_\mu \frac{e^{-\beta \norm{\boldsymbol{\xi_\mu - x_n}}^2}}{ z_\mu }$ and $z_\mu = \sum_m e^{-\beta \norm{\boldsymbol{\xi_\mu - x_m}}^2}$.
\subsection{In 1D}
Let's consider the case of two phenotypes, $\xi_1$ and $\xi_2$. As the dynamics rely on euclidean distances, this means there exists a stable subspace of at most one dimension (generally, $(N_{phenotypes} - 1)$ dimensions). Hence, we write
\begin{subequations}
    \begin{gather}
    \xi_1 = -\frac{d}{2}, \quad \xi_2 = +\frac{d}{2}, \\
    \boldsymbol{x}_n \xrightarrow{} x_n,
\end{gather}
\end{subequations}
keeping track only of the dynamics within this stable subspace, with no orthogonal component in $\boldsymbol{x}_n$ by i.c. (although more generally, orthogonal directions will decay rapidly). The dynamics simplify into, 
\begin{equation}
    \partial_t x_n = \frac{1}{Z_n} \Big(
        e^{-\beta (x_n - \frac{d}{2})^2} \frac{1}{z_2}
        - e^{-\beta (x_n + \frac{d}{2})^2} \frac{1}{z_1} \Big) \frac{d}{2} - x_n,
\end{equation}
with $Z_n = e^{-\beta (x_n - \frac{d}{2})^2} \frac{1}{z_2} + e^{-\beta (x_n + \frac{d}{2})^2} \frac{1}{z_1}$. Altogether,
\begin{gather} \label{eq: S 1D ISING initial}
    \partial_t x_n = \Big(
        \frac{e^{-\beta (x_n - \frac{d}{2})^2} - e^{-\beta (x_n + \frac{d}{2})^2} \frac{z_2}{z_1}}
        { e^{-\beta (x_n - \frac{d}{2})^2}  + e^{-\beta (x_n + \frac{d}{2})^2} \frac{z_2}{z_1}} \Big) \frac{d}{2} - x_n
\end{gather}
where $\frac{z_2}{z_1}$ is effectively the only free mean-field parameter. 
Moving $\frac{z_2}{z_1}$ to the exponent, and balancing out the terms, we obtain
\begin{gather}
    \partial_t x_n = \Big(
        \frac{e^{-\beta (x_n - \frac{d}{2})^2 - \frac{1}{2}\ln(\frac{z_2}{z_1})} - e^{-\beta (x_n + \frac{d}{2})^2 + \frac{1}{2}\ln(\frac{z_2}{z_1})} }
        { e^{-\beta (x_n - \frac{d}{2})^2 - \frac{1}{2}\ln(\frac{z_2}{z_1})}  + e^{-\beta (x_n + \frac{d}{2})^2 + \frac{1}{2}\ln(\frac{z_2}{z_1})} } \Big) \frac{d}{2} - x_n,
\end{gather}
which when cancelling out common terms (i.e. $e^{-\beta (x^2_n + \frac{d^2}{4})}$) leads to
\begin{gather}
    \partial_t x_n = \Big(
        \frac{e^{  (\beta x_n d  - \frac{1}{2}\ln(\frac{z_2}{z_1}))} - e^{- (\beta x_n d - \frac{1}{2}\ln(\frac{z_2}{z_1}))} }
        { e^{  (\beta x_n d  - \frac{1}{2}\ln(\frac{z_2}{z_1}))}  + e^{- (\beta x_n d - \frac{1}{2}\ln(\frac{z_2}{z_1}))} } \Big) \frac{d}{2} - x_n,
\end{gather}
which is fits the definition of a hyperbolic tangent;
\begin{gather} \label{eq: S 1d Ising final}
    \boxed{\partial_t x_n = 
        \tanh \Big( \beta x_n d - \frac{1}{2} \ln(\frac{z_2}{z_1}) \Big) \frac{d}{2} - x_n.}
\end{gather}
whose fixed points are akin to the magnetization of the mean-field Ising.
\subsection{In 2D}
A similar derivation can be made with three phenotypes. Again, taking advantage of our stable 2D subspace, we write
\begin{subequations}
    \begin{gather}
    \boldsymbol{\xi}_1 = (-\frac{d_x}{2}, -\frac{d_y}{2}), \quad \boldsymbol{\xi}_2 = (\frac{d_x}{2}, -\frac{d_y}{2}), \quad \boldsymbol{\xi}_3 = (\delta_x, \frac{d_y}{2}), \\
    \boldsymbol{x}_n \xrightarrow{} (x_n, y_n).
\end{gather}
\end{subequations}
Importantly, the Ising formulation requires no special symmetry, and any 3-phenotype geometry can be described by the degrees of freedom $d_x$, $d_y$, and $\delta_x$ (up to a global rotation, or translation which will not affect the dynamics). Starting with the $y$-dynamics, we can write
\begin{gather} \label{eq: S y-dynamics initial}
    \partial_t y_n = \frac{1}{Z_n} \Big(  \frac{1}{z_3} e^{-\beta (x_n - \delta_x)^2} e^{-\beta (y_n - \frac{d_y}{2})^2} - (\frac{1}{z_2} e^{-\beta (x_n - \frac{d_x}{2})^2} + \frac{1}{z_1}e^{-\beta (x_n + \frac{d_x}{2})^2})e^{-\beta (y_n + \frac{d_y}{2})^2}  \Big) \frac{d_y}{2} - y_n
\end{gather}
for simplicity, we define placeholder variables $q_3 = \frac{1}{z_3} e^{-\beta (x - \delta_x)^2}$ and $q_2 = \frac{1}{z_2} e^{-\beta (x_n - \frac{d_x}{2})^2} + \frac{1}{z_1}e^{-\beta (x_n + \frac{d_x}{2})^2}$, to write
\begin{gather}
    \partial_t y_n = \Big(  \frac{ e^{-\beta (y_n - \frac{d_y}{2})^2} -  e^{-\beta (y_n + \frac{d_y}{2})^2} \frac{q_2}{q_3}}{ e^{-\beta (y_n - \frac{d_y}{2})^2} +  e^{-\beta (y_n + \frac{d_y}{2})^2}\frac{q_2}{q_3}}  \Big) \frac{d_y}{2} - y_n.
\end{gather}
The above is directly related to eq.~\eqref{eq: S 1D ISING initial}, allowing us to write 
\begin{gather}
    \partial_t y_n = \tanh \Big( \beta y_n d_y - \frac{1}{2} \ln(\frac{q_2}{q_3}) \Big) \frac{d_y}{2} - y_n
\end{gather}
which can be simplified further by decomposing at $\ln(\frac{q_2}{q_3})$ into a mean-field portion and a unit-specific portion,
\begin{gather}
    \boxed{
        \partial_t y_n = \tanh \Big( \beta y_n d_y - \frac{1}{2} \ln(\frac{z_3}{z_1}) - \phi(x_n, z_1/z_2) \Big) \frac{d_y}{2} - y_n,
    }
\end{gather}
where 
\begin{gather}
    \phi(x_n, z_1/z_2) = \frac{1}{2}\ln \Big( \frac{e^{-\beta (x_n - \frac{d_x}{2})^2}\frac{z_1}{z_2} + e^{-\beta (x_n + \frac{d_x}{2})^2}}{ e^{-\beta (x_n - \delta_x)^2}} \Big).
\end{gather}
For the $x_n$, we write
\begin{gather}
    \partial_t x_n = \frac{1}{Z_n} \Big( \frac{1}{z_2} e^{-\beta (x_n - \frac{d_x}{2})^2} - \frac{1}{z_1} e^{-\beta (x_n + \frac{d_x}{2})^2} \Big) e^{-\beta (y_n + \frac{d_y}{2})^2} \frac{d_x}{2} + \frac{1}{Z_n} \Big( \frac{1}{z_3} e^{-\beta (x_n - \delta_x)^2} e^{-\beta (y_n - \frac{d_y}{2})^2} \Big) \delta_x - x_n,
\end{gather}
which implicitly mixes in $y$-dynamics due to the $\delta_x$ asymmetry. Indeed we can write the $y$-dynamics of eq.~\eqref{eq: S y-dynamics initial} as effectively $\partial_t y_n = \frac{A - B}{A + B} \frac{d_y}{2} - y_n$ with $A = \frac{1}{z_3} e^{-\beta (x_n - \delta_x)^2} e^{-\beta (y_n - \frac{d_y}{2})^2}$, moving terms around we have $(\partial_t y_n + \frac{d_y}{2})\frac{\delta_x}{d_y} = \frac{A}{A+B} \delta_x - \frac{\delta_x}{d_y} y_n$. Making this explicit, we have
\begin{gather}
    \partial_t \big(x_n - \frac{\delta_x}{d_y} (y_n + \frac{d_y}{2})\big) = \frac{1}{Z_n} \Big( \frac{1}{z_2} e^{-\beta (x_n - \frac{d_x}{2})^2} - \frac{1}{z_1} e^{-\beta (x_n + \frac{d_x}{2})^2} \Big) e^{-\beta (y_n + \frac{d_y}{2})^2} \frac{d_x}{2} - \big(x_n-\frac{\delta_x}{d_y}(y_n + \frac{d_y}{2})\big).
\end{gather}
This is hyperbolic-like, albeit with artifacts due to the 2-dimensional nature of this system. Looking closer at the first term, 
\begin{subequations}
    \begin{align}
        & \ \frac{1}{Z_n} \Big( \frac{1}{z_2} e^{-\beta (x_n - \frac{d_x}{2})^2} - \frac{1}{z_1} e^{-\beta (x_n + \frac{d_x}{2})^2} \Big) e^{-\beta (y_n + \frac{d_y}{2})^2}\\[10pt]
        & \quad = \frac{\Big( \frac{1}{z_2} e^{-\beta (x_n - \frac{d_x}{2})^2} - \frac{1}{z_1} e^{-\beta (x_n + \frac{d_x}{2})^2} \Big) e^{-\beta (y_n + \frac{d_y}{2})^2}}{\Big( \frac{1}{z_2} e^{-\beta (x_n - \frac{d_x}{2})^2} + \frac{1}{z_1} e^{-\beta (x_n + \frac{d_x}{2})^2} \Big) e^{-\beta (y_n + \frac{d_y}{2})^2} + \frac{1}{z_3} e^{-\beta (x_n - \delta_x)^2} e^{-\beta (y_n - \frac{d_y}{2})^2}}\\[10pt]
        & \hspace{2cm} = k_n \frac{ \frac{1}{z_2} e^{-\beta (x_n - \frac{d_x}{2})^2} - \frac{1}{z_1} e^{-\beta (x_n + \frac{d_x}{2})^2} }{ \frac{1}{z_2} e^{-\beta (x_n - \frac{d_x}{2})^2} + \frac{1}{z_1} e^{-\beta (x_n + \frac{d_x}{2})^2} }
    \end{align}
    \begin{equation}
        k_n = \frac{1}{1 + \frac{\frac{1}{z_3}e^{-\beta (x_n - \delta_x)^2} e^{-\beta (y_n - \frac{d_y}{2})^2}}{\Big( \frac{1}{z_2} e^{-\beta (x_n - \frac{d_x}{2})^2} + \frac{1}{z_1} e^{-\beta (x_n + \frac{d_x}{2})^2} \Big) e^{-\beta (y_n + \frac{d_y}{2})^2}}}.
    \end{equation}
\end{subequations}
Thus, we have
\begin{gather}
    \partial_t \big(x_n - \frac{\delta_x}{d_y} (y_n + \frac{d_y}{2})\big) = k_n 
    \frac{ \frac{z_1}{z_2} e^{-\beta (x_n - \frac{d_x}{2})^2} -  e^{-\beta (x_n + \frac{d_x}{2})^2} }{ \frac{z_1}{z_2} e^{-\beta (x_n - \frac{d_x}{2})^2} +  e^{-\beta (x_n + \frac{d_x}{2})^2} } \frac{d_x}{2} - \big(x_n-\frac{\delta_x}{d_y}(y_n + \frac{d_y}{2})\big),
\end{gather}
whose first term is as in eq.~\eqref{eq: S 1D ISING initial}, and leads to
\begin{gather} \label{eq: S x-dynamics final}
    \boxed{
        \partial_t \big(x_n - \frac{\delta_x}{d_y} (y_n + \frac{d_y}{2})\big) = k_n \tanh\Big( \beta x_n d_x - \frac{1}{2}\ln(\frac{z_2}{z_1}) \Big) \frac{d_x}{2} - \big(x_n-\frac{\delta_x}{d_y}(y_n + \frac{d_y}{2})\big).
    }
\end{gather}
\subsubsection{The isosceles case}
The above equation hints at a change of $x$-variable, due to the asymmetry in $\xi_3$. Importantly, the isosceles case, where $\delta_x = 0$, considerably simplifies eq~\eqref{eq: S x-dynamics final} into,
\begin{gather}
        \partial_t x_n = k_n \tanh\Big( \beta x_n d_x - \frac{1}{2}\ln(\frac{z_2}{z_1}) \Big) \frac{d_x}{2} - x_n.
\end{gather}
which is only coupled to $y_n$ through $k_n$ (effectively weakening $\pm\frac{d_x}{2}$ attraction for cells near $\xi_3$). 
\section{The differentiation timescale} \label{section: Supp Timescale}
In this section, we detail the derivations relating to differentiation timescales and progenitor locations, and identify $\beta$ as the main factor that directs the system towards low-dimensional, sequential, differentiation.  
\subsection{The piecewise approximation in 1D} \label{section: Supp piecewise Ising}
Recall the dynamics of the 1D system in the Ising formulation (eq.~\ref{eq: S 1d Ising final}),
\begin{gather*}
    \partial_t x_n = 
        \tanh \Big( \beta x_n d - \frac{1}{2} \ln(\frac{z_2}{z_1}) \Big) \frac{d}{2} - x_n
\end{gather*}
taking a purely linear approximation is enough to describe annihilation bifurcation of phenotypes, however coarsens away important features of dynamics. Instead, we approximate $\tanh$ by pieces
\begin{gather}
    \tanh(x) \approx \begin{cases}
        x, & \quad \vert x \vert \leq 1 \\
        +1, & \quad x > 1\\
        -1, & \quad x < 1
    \end{cases}
\end{gather}
thus incorporating saturation. The piecewise dynamics are then,
\begin{gather}
    \partial_t x_n \approx \begin{cases}
        \Big( \beta x_n d - \frac{1}{2} \ln(\frac{z_2}{z_1}) \Big) \frac{d}{2} - x_n, 
        & \quad \frac{\frac{1}{2}\ln(\frac{z_2}{z_1}) - 1}{\beta d} \leq x_n \leq \frac{\frac{1}{2}\ln(\frac{z_2}{z_1}) + 1}{\beta d} \\
        +\frac{d}{2} - x_n, & \quad x_n > \frac{\frac{1}{2}\ln(\frac{z_2}{z_1}) + 1}{\beta d}\\
        -\frac{d}{2} - x_n, & \quad x_n < \frac{\frac{1}{2}\ln(\frac{z_2}{z_1}) - 1}{\beta d}.
    \end{cases}
\end{gather}
This implies two regimes, an exponential explosion (for $\beta \geq \frac{2}{d^2}$) described by the first piece, and an exponential relaxation given by the last two pieces. For simplicity (w.l.o.g.), let's assume our dynamics begin in the first regime, $\frac{\frac{1}{2}\ln(\frac{z_2}{z_1}) - 1}{\beta d} \leq x_n(0) \leq \frac{\frac{1}{2}\ln(\frac{z_2}{z_1}) + 1}{\beta d}$. We can then integrate our dynamics to write
\begin{gather} \label{eq: S exponential explosion}
    x_n(t) = \Big( x_n(0) + \frac{d}{2}\int_0^t \ln(\frac{z_2(t')}{z_1(t')}) e^{-(\frac{1}{2}\beta d^2 - 1)t'}dt'\Big) e^{(\frac{1}{2}\beta d^2 - 1)t}, \quad t \leq t^*_n
\end{gather}
where $t^*_n$ defines the time at which the unit leaves the explosion regime (i.e. goes beyond the bounds of the first piece). Notice that the mean-field variables can be time-changing, and as such we must integrate $\int_0^t \ln(\frac{z_2(t')}{z_1(t')}) e^{-(\frac{1}{2}\beta d^2 - 1)t'}dt'$ over the time-changing impulse it generates. Outside the exponential explosion regime, the relaxation equation(s) integrate to 
\begin{gather} \label{eq: S exponential relaxation}
    x_n(t) = \Big( x(t^*_n) \mp \frac{d}{2} \Big)e^{-(t-t^*)} \pm \frac{d}{2}, \quad t \geq t^*_n,
\end{gather}
where $\pm$ corresponds to the second and third pieces of our piecewise function.
\subsection{The differentiation timescale in 1D}
In general, units first converge to the central saddle before any differentiation takes place. This is driven by $\frac{z_2}{z_1}$, as imbalances in signaling/occupancy tilt the landscape, biasing the dynamics against the over-populated phenotype, until undifferentiated units are at a state which perfectly balances the signals (i.e. the geometric mean / the central saddle). Mathematically, this is because differentiation dynamics are a first-order perturbation and are only non-negligibly small when the zero-th order (collective convergence) disappears (i.e. at a fixed point)

For simplicity, let's focus our analysis on the second part of the dynamics, assuming units have have been initialized (or equivalently have already converged) to the central saddle (up to some stochasticity), and are about to differentiate. Additionally, let us assume mean-field signals $\frac{z_2}{z_2} \approx 1$ remain balanced throughout differentiation. Then, we can define
\begin{gather}
    \langle \vert x_n(\tau^\epsilon) \vert \rangle = \epsilon
\end{gather}
the time $\tau^\epsilon$ units take to move (on average) $\epsilon$ away from the central saddle (the origin, by construction). Which by symmetry, is effectively the time units take to be $\sim 2\epsilon$-differentiated. When $\epsilon$ is chosen to be within the exponential explosion regime ($\epsilon \leq \frac{1}{\beta d}$), we can write using eq.~\eqref{eq: S exponential explosion}
\begin{gather} \label{eq: tau epsilon 1d}
    \tau^\epsilon = \frac{1}{\frac{1}{2}\beta d^2 - 1} \ln \Big(\frac{\epsilon}{\langle \vert x_n(0) \vert \rangle} \Big), \quad \epsilon \leq \frac{1}{\beta d}.
\end{gather}
Holding $\epsilon$, $x_n(0)$ and $d$ fixed, we see that the time to differentiate in 1D explodes as we decrease $\beta$, with an asymptote at $\beta = \frac{2}{d^2}$ leading to an infinite dwell-time, as is common for bifurcations. More generally, we can define for larger $\epsilon$
\begin{gather}\label{eq: S t star}
    \tau^\epsilon\Big\vert_{\epsilon=\frac{1}{\beta d}} = t^* = \frac{1}{\frac{1}{2}\beta d^2 - 1} \ln \Big( \frac{1}{\beta d \vert \langle x_n(0) \rangle \vert} \Big), 
\end{gather}
while for $\epsilon > \frac{1}{\beta d}$ we must use eq.~\eqref{eq: S exponential relaxation} to obtain
\begin{gather} \label{eq: tau epsilon 1d relaxation}
    \tau^\epsilon = t^* + \ln \Big( \frac{ \frac{d}{2} - \frac{1}{\beta d}}{\frac{d}{2} - \epsilon} \Big), \quad \epsilon > \frac{1}{\beta d}.
\end{gather}
\subsection{The differentiation timescale in 2D} 
In 2D, things are slightly more complicated for two reasons. First, our dimensions are coupled by non-trivial parameters/variables (i.e. $k_n$ and $\phi(x_n, z_1/z_2)$) which are difficult to disentangle. Second, $\tau$ differences (or ratios) are an incomplete measure of separation due to the different time- and length-scales, and require additional context.  
\subsubsection{Static landscape approximation}
For the first, let's consider a simplification of the ($\delta_x = 0$) system 
\begin{itemize}
    \item we assume, as before that $z_1/z_2$ are balanced (either through initial conditions, or unit convergence pre-differentiation
    \item we ignore the coupling between $x_n$ and $y_n$ (i.e. set $k_n = 1$), treating our estimates on $\tau^\epsilon_x$ as effective upper bounds (as $k_n \leq 1$ would otherwise slow down $x_n$ differentiation)
    \item we consider geometries where $d_y > \frac{\sqrt{3}}{2}d_x$, such that $y$-differentiation occurs first, and we focus on timepoints s.t. $\vert x_n(t) \vert \leq \epsilon \ll 1$ allowing us to reduce $\phi(x_n, z_1/z_2) \xrightarrow{} -\frac{1}{2} \ln(2)$ and $\ln(\frac{z_1}{z_3}) \approx \ln{\frac{\sum_m e^{-\beta(y_m+d_y/2)^2}}{\sum_m e^{-\beta(y_m-d_y/2)^2}}}$ as the $x_n$ portion in $\phi$ and $\frac{1}{2}\ln{\frac{z_1}{z_3}}$ cancel for low $\epsilon$.   
\end{itemize}
This completely disentangles our $x_n$ and $y_n$ equations into,
\begin{subequations}
    \begin{gather}
        \partial_t x_n = \tanh \Big( \beta x_n d_x \Big) \frac{d_x}{2} - x_n, \\
        \partial_t y_n = \tanh \Big( \beta y_n d_y + \frac{1}{2}\ln(2) - \frac{1}{2} \ln(\frac{z_{+d_y/2}}{z_{-d_y/2}}) \Big) \frac{d_y}{2} - y_n
    \end{gather}
\end{subequations}
\subsubsection{The effective progenitor location}
Recall that, typically, units first move towards (or are initialized at) a central/first saddle, differentiate along the $y$, then move towards a second saddle before differentiating along the $x$. 
Hence, a more intuitive measure of how `separated' $x$ and $y$ differentiation are, is the seperation between these two differentiation points. Well separated branching points lead to the characteristically low-dimensional dynamics --- differentiation occurs only along one axis at a time ---
while, more adjacent branching points lead to trajectories that vary (non-negligibly) across both $x$ and $y$, and in the limit to trifurcations.
More precisely, it suffices to calculate the position along the fast axis ($\langle \vert y_n \vert \rangle$) when units are on average $2\epsilon$-differentiated along the slow axis ($x$).
First, we calculate the time to $2\epsilon$-differentiation along the slow axis using eq.~\eqref{eq: tau epsilon 1d}-\eqref{eq: tau epsilon 1d relaxation}
\begin{gather}
    \tau^\epsilon_x = \begin{cases}
        \frac{1}{\frac{1}{2}\beta d_x^2 - 1} \ln \Big(\frac{\epsilon}{\langle \vert x_n(0) \vert \rangle} \Big), & \quad \beta \leq \frac{1}{\epsilon d} \\[10pt]
        t^*_x + \ln \Big( \frac{ \frac{d_x}{2} - \frac{1}{\beta d_x}}{\frac{d_x}{2} - \epsilon} \Big), & \quad \beta > \frac{1}{\epsilon d_x}
    \end{cases}
\end{gather}
where for high $\beta$ $x$-differentiation will enter the exponential relaxation regime, which we must account for. Following our phenotype geometry $d_y > \frac{\sqrt{3}}{2} d_x$, we assume $y$-dynamics very quickly enter the exponential relaxation regime (s.t. $\tau^\epsilon_x \gg t^*_{y}$). Hence, using eq.~\eqref{eq: S exponential relaxation} with $\langle y_n(t^*_y \approx 0) \rangle \approx -\frac{1}{6} d_y$ (i.e. the geometric mean of the $\boldsymbol{\xi}$'s)
\begin{gather}
    \langle y_n(t) \rangle \approx \frac{d_y}{3}e^{-t} - \frac{d_y}{2},
\end{gather}
which, when combined with $\tau^\epsilon_x$ leads to
\begin{subequations}
    \begin{gather}
        \boxed{
            \langle y_n(t) \rangle \approx \frac{d_y}{3} e^{-\frac{1}{\frac{1}{2}\beta d_x^2 - 1} \ln \Big(\frac{\epsilon}{\langle \vert x_n(0) \vert \rangle} \Big)} - \frac{d_y}{2},  \quad \beta \leq \frac{1}{\epsilon d_x}
        } \\
        \label{eq: double exponential explosion lengthscale}
        \boxed{
        \langle y_n(t) \rangle \approx \frac{d_y}{3} e^{-\frac{1}{\frac{1}{2}\beta d_x^2 - 1} \ln \Big(\frac{1}{\beta d_x \langle \vert x_n(0) \vert \rangle} \Big) - \ln \Big( \frac{ \frac{d_x}{2} - \frac{1}{\beta d_x}}{\frac{d_x}{2} - \epsilon} \Big)} - \frac{d_y}{2},  \quad \beta > \frac{1}{\epsilon d_x}
        }
    \end{gather}
\end{subequations}
where $t^*_x$ was substituted in with eq.~\eqref{eq: S t star}. Note that eq.~\eqref{eq: double exponential explosion lengthscale} also holds as a good approximation for $\beta \leq \frac{1}{\epsilon d_x}$.

\clearpage
\section{Simulation parameters for all figures}
\begin{table*}[ht]
    \caption{\textbf{Value of model and simulation parameters.} The table lists parameter values for core simulations.}
    \label{tab:parameters}
    \begin{ruledtabular}
    \begin{tabular}{llccccccc}
    \multirow{2}{*}{Type} & \multirow{2}{*}{Param.} & \multicolumn{7}{c}{Figure} \\
    \cline{3-9}
    & & 1B--C & 1D--E & 2A--D & 3C, E, F & 4B & 4D--E & 5A  \\
    \hline
    \multirow{2}{*}{\rotatebox[origin=c]{0}{$\xi$}} 
    & $N_\xi$      & 3 & 600 ($60\times 10$) & 3 & 3 & 3 & 5 & Data  \\
    & State dim.   & 2 & 784 & 2 & 2 & 2 & 2 & PCA  \\
    \hline
    \multirow{3}{*}{\rotatebox[origin=c]{90}{Pop.}}
    & $K$          & 500 & 400 & 100 & 500 & $8{\times}10^4$ & $2{\times}10^5$ & $5{\times}10^5$  \\
    & $t_{\max}$   & 1000 & 3000 & 800 & 1500 & $2.4{\times}10^6$ & 66666 & 10000 \\
    & Growth $a$   & -- & -- & -- & -- & 1/30 & 3 & 50 \\
    \hline
    \multirow{5}{*}{\rotatebox[origin=c]{90}{Dyn.}}
    & $\beta$      & 1.5 & 0.36 & $\{1.5, 3, 11\}$\footnote{Fig. 2D sweeps $\beta$ from 1.5 to 13.0.} & 1.5 & 1.5 & 1.0 & 0.025 \\
    & $\gamma$     & $0$ & $10^{-5}$ & 0 & $0$ & $\boldsymbol{\gamma}_0$\footnote{For Fig. 4B, $\boldsymbol{\gamma}_0 = (3200, 600, 4)$ to show the instability of the high $\Tilde{z}_2$ choice-like geometry.} & $10^{-2}$ & $10^{-5}$ \\
    & lr           & 0.014 & 0.0051 & 0.03 & 0.014 & 0.02 & 0.02 & 0.06 \\
    & Noise\footnote{Order of magnitude, e.g. $6.0 \xrightarrow[]{} 10^{-6}$}        & 6.0\footnote{Noise present only in the initial conditions.} & 6.0 & 7 & 6.0\footnotemark[4] & 4.0 & 2.75 & 2.0\\
    & Seed                  & 41 & 42 & 42 & 41 & 4 & 41 & 42 \\
    \hline
    \multirow{2}{*}{\rotatebox[origin=c]{90}{Pert.}}
    & $t_{\text{perturb}}$ & -- & -- & -- & 700 & -- & -- & -- \\
    & Death win.           & -- & -- & -- & 50 & -- & -- & -- \\
    \end{tabular}
    \end{ruledtabular}
\end{table*}

\section{Numerical Implementation and Reproducibility}
Simulations were implemented in JAX \cite{jax} and run on a single GPU (NVIDIA RTX 5000 Ada). Main simulations typically complete in a few minutes, while the more demanding catastrophe space calculations run in tens of minutes (with the possibility to parallelize over multiple GPUs for faster computation). In general, all code can also be run on CPU-only hardware, however highly parallelized calculations become significantly slower (e.g. catastrophe calculations, and energy manifold calculations). Two-dimensional figures were built with Matplotlib \cite{matplotlib}, while most three-dimensional panels were generated using vedo \cite{vedo}, a library more suited to 3D visualization. All code is publicly available at \href{github.com/nacer-eb/Sandscapes}{github.com/nacer-eb/Sandscapes}.

\section{Preprocessing steps for single cell RNA-seq data}
\paragraph{Data source}
The single-cell RNA data is obtained from the hematopoietic lineage-tracing dataset \cite{Weinreb2020} using GEO accession number \textit{GSE140802} - a script to automatically download and extract the dataset is provided. The dataset includes  normalized expression counts, gene names, cell barcodes, per-cell metadata (time point and cell-type annotation), and the clonal barcode matrix linking cells to lentivirally labeled clonal families. Note that only the  in-vitro portion of the dataset is used. 
\paragraph{Cell annotation}
Cells were filtered by time point and type, to keep only 11 broad cell-types. Namely, Undifferentiated cells from Day-2, and 10 distinct terminally differentiated cell types from Day-6.
\paragraph{Gene filtering}
Mitochondrial (MT-) and ribosomal (RPS, RPL) genes were removed, along with a curated set of genes known to reflect technical noise or other such artifacts.
\paragraph{Dataset balancing}
Cell type populations are unbalanced in the original dataset with some cell types much more highly represented than others. To avoid bias from an uneven cell type distribution, a balanced subset is constructed by randomly sampling 200 cells within each of the 11 pre-selected cell types. This sampling can be performed with a random seed for reproducibility.
\paragraph{Preprocessing}
Gene expression was normalized using the Pearson residuals method, under a regularized negative binomial model. For each gene, an overdispersion parameter $\theta$ was estimated via method-of-moments and regularized, after which residuals were computed and clipped at $\sqrt{N_{cells}}$ to avoid outliers having an oversized effect or dominating the entire dataset. This method is akin to the one implemented in scanpy and described in \cite{Lause2021}.
\paragraph{Feature selection and dimensional reduction}
Genes were ranked by residual variance, and the top 2,000 most variable genes were retained in addition to a curated set of progenitor marker genes known from literature (e.g. \textit{Kit}, \textit{Cd34}, ...) to be used in the analysis. PCA was applied to the resulting expression matrix (2200 cells $\times$ 2006 genes) keeping only 100 principal components. A supervised (3d) NCA embedding is also computed using the 100 PCs, for visualization purposes only.

\section{Supplemental Figures}
\begin{figure}[h]
    \centering
    \includegraphics[width=0.99\linewidth]{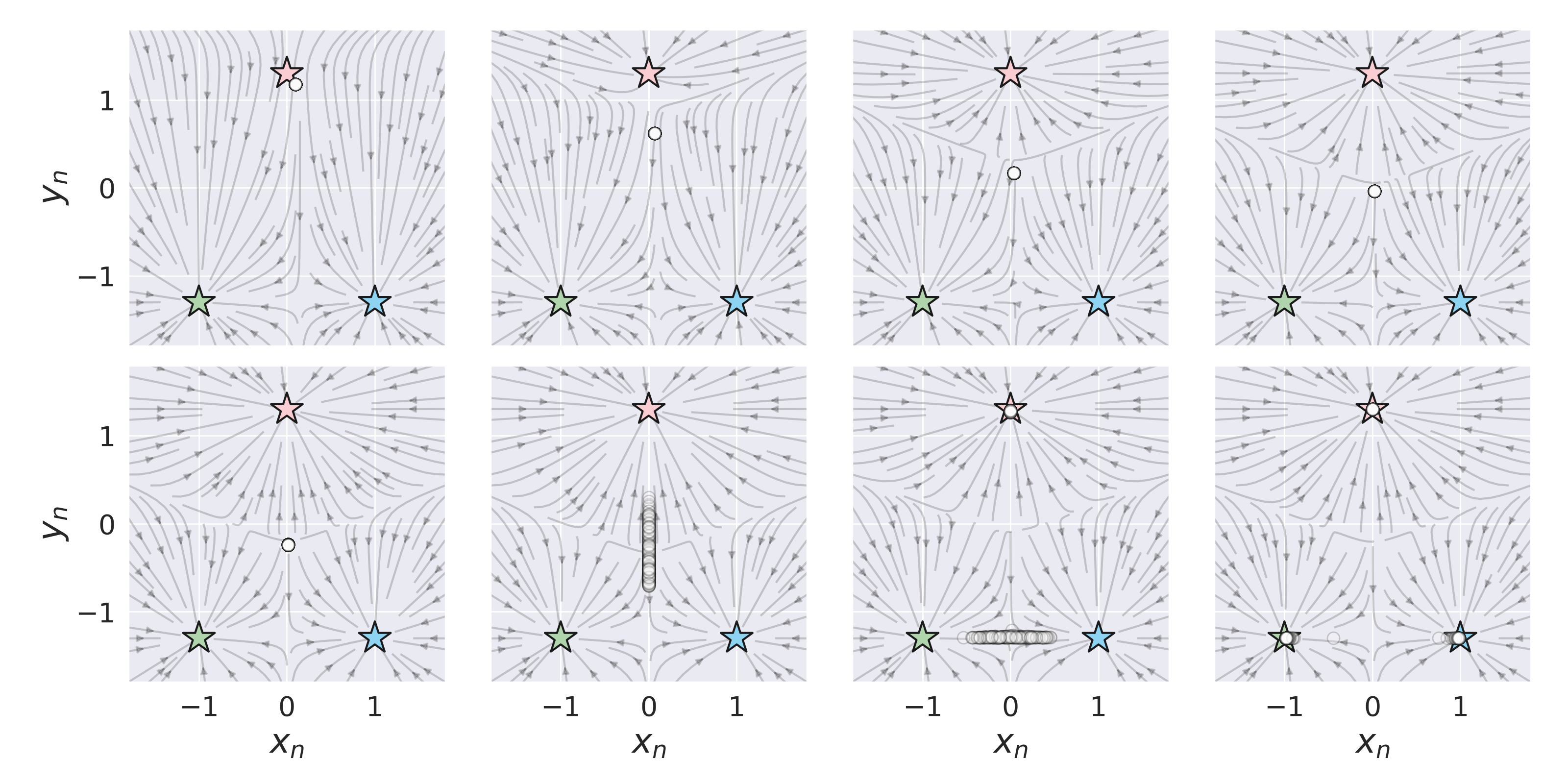}
    \caption[State space flow during differentiation, for the 3-phenotype system.]{Dynamical flow lines (streamplot) for the 3-phenotype system at various dynamical times, for fixed $\beta = 1.5$ and phenotype geometry $d_x = 2.0$ and $d_y=2.6$. Phenotypes $\boldsymbol{\xi}$ are drawn as stars, while units are drawn as black/white points. This serves as complementary to Fig.~\ref{fig:fig1}B, C,. }
    \label{fig:SFig-dynamical-flow}
\end{figure}

\begin{figure}[h]
    \centering
    \includegraphics[width=0.99\linewidth]{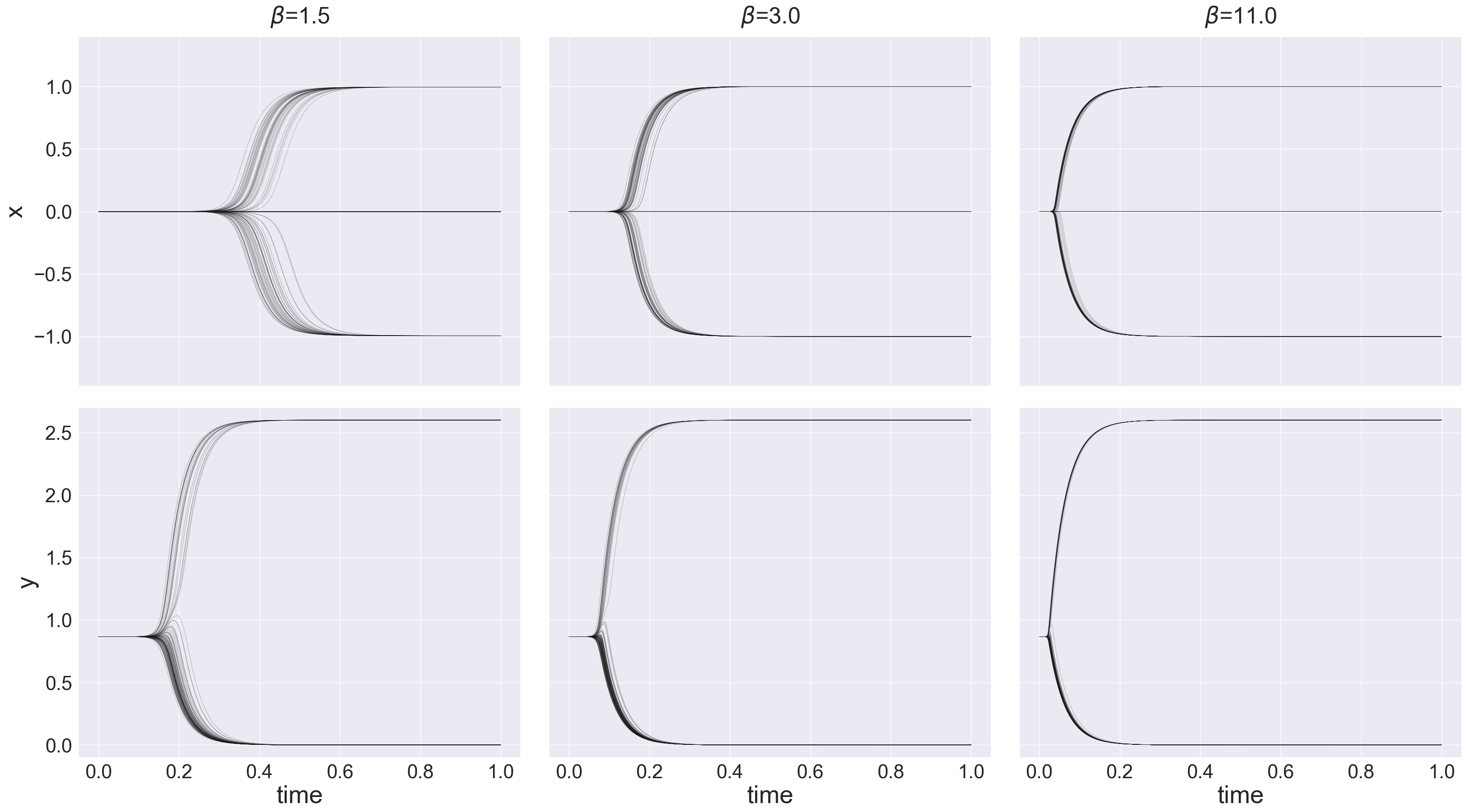}
    \caption[Dynamical traces of the 3-phenotype system.]{Dynamical traces $x_n(t)$ and $y_n(t)$ of the 3-phenotype system, for various $\beta$ values ($1.5$, $3.0$, $11.0$), for the same simulations as Fig.~\ref{fig:fig2}. Phenotype geometry is fixed with $d_x = 2.0$ and $d_y=2.6$.}
    \label{fig:SFig1}
\end{figure}

\begin{figure}[h]
    \centering
    \includegraphics[width=0.99\linewidth]{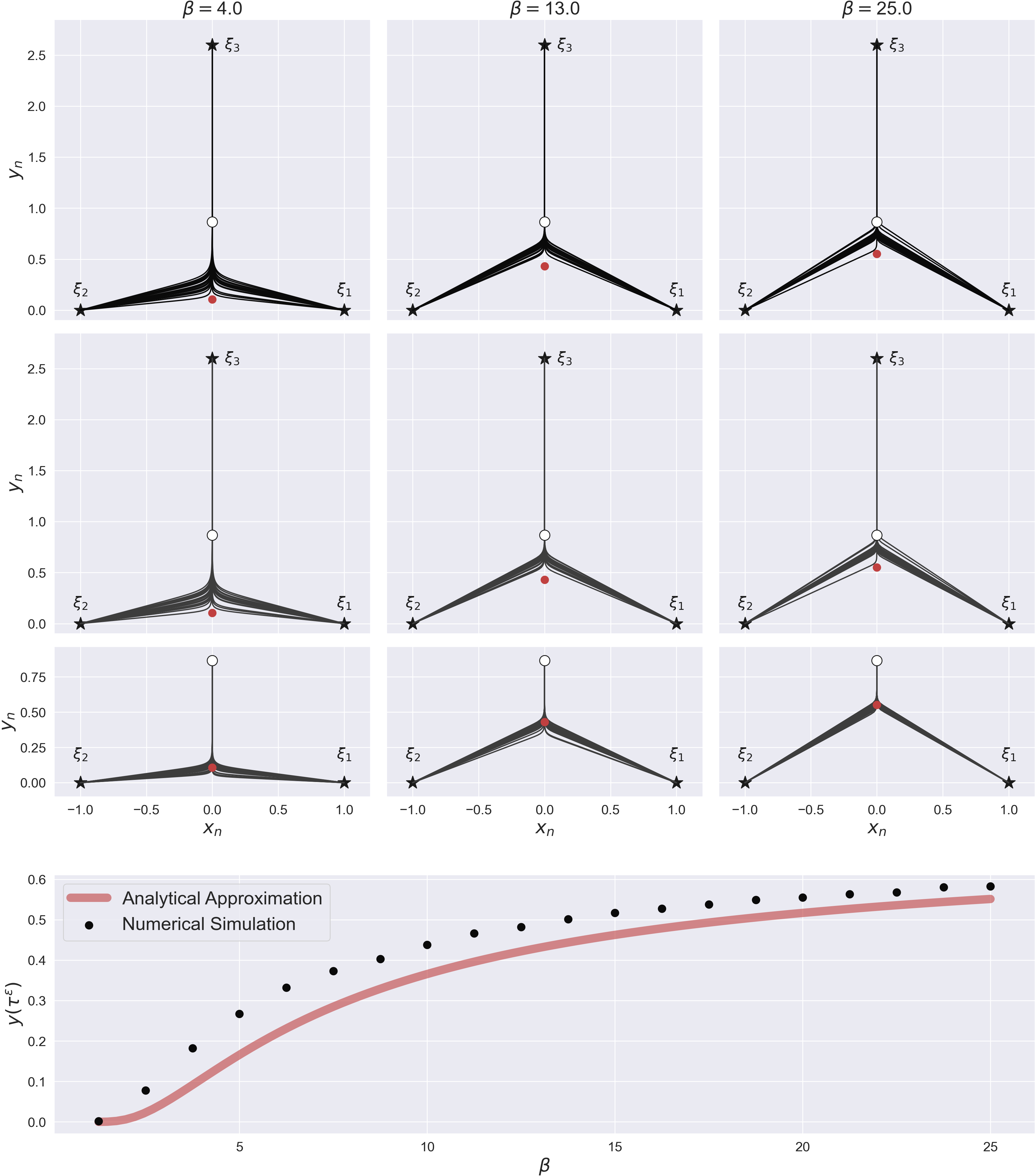}
    \caption[Analytical approximation of the progenitor location, for $d_y=2.6$.]{(First row) Dynamical trajectories $x_n(t)$ and $y_n(t)$ of the 3-phenotype system, for various $\beta$ values ($4.0$, $13.0$, $25.0$). (Second row) Dynamical trajectories $x_n(t)$ and $y_n(t)$ of the 3-phenotype system, with the Ising formulation. (Third row) Dynamical trajectories $x_n(t)$ and $y_n(t)$ of the 3-phenotype system, with the \textit{uncoupled} Ising formulation. (Bottom) Numerical estimation of the progenitor location using the full Ising sysem (black), compared to the analytical approximation (red) derived in \S\ref{section: Supp Timescale}. Phenotype geometry is fixed with $d_x = 2.0$ and $d_y=2.6$.}
    \label{fig:SFig-progenitor-1}
\end{figure}

\begin{figure}[h]
    \centering
    \includegraphics[width=0.99\linewidth]{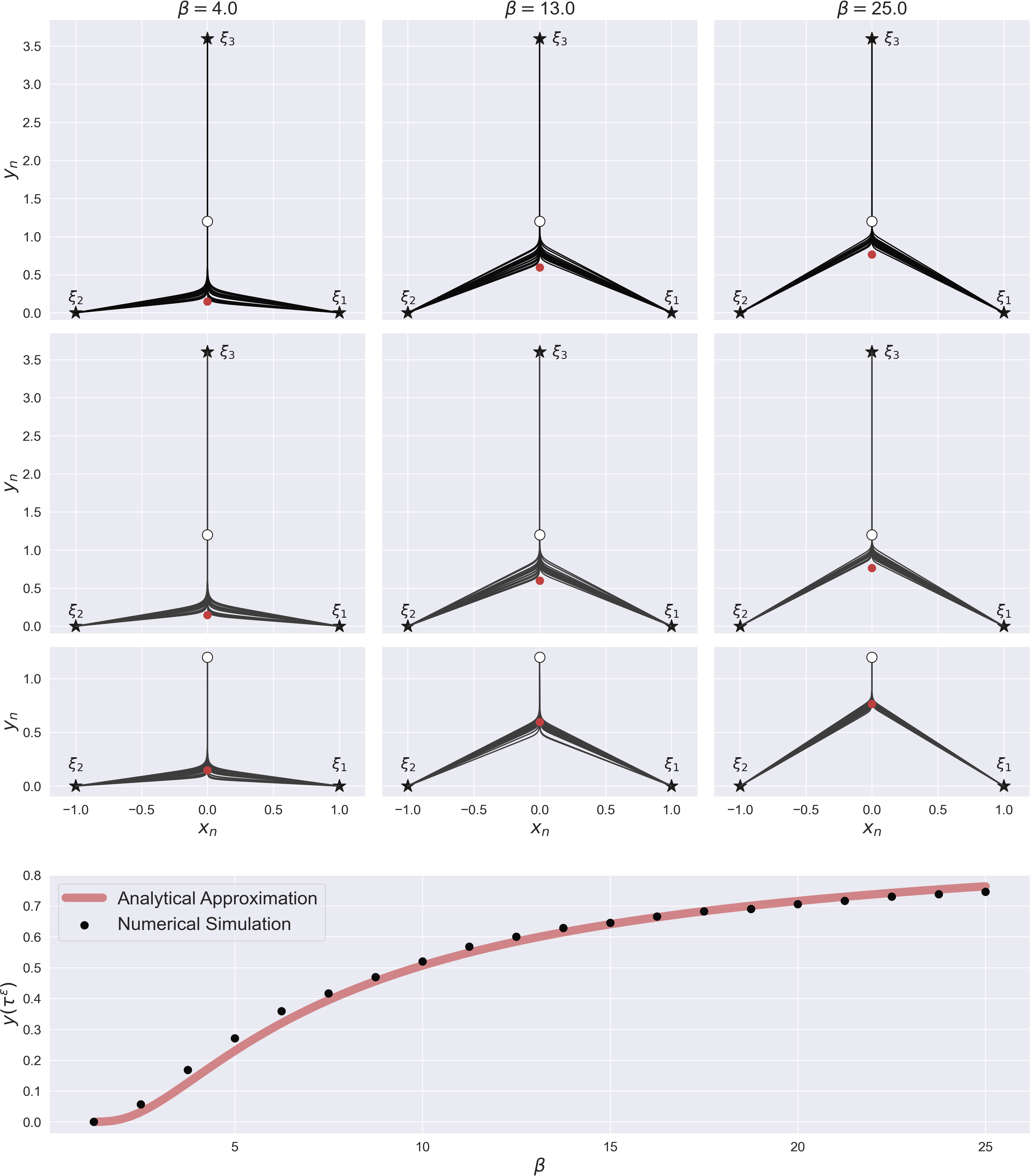}
    \caption[Analytical approximation of the progenitor location, for $d_y=3.6$.]{(First row) Dynamical trajectories $x_n(t)$ and $y_n(t)$ of the 3-phenotype system, for various $\beta$ values ($4.0$, $13.0$, $25.0$). (Second row) Dynamical trajectories $x_n(t)$ and $y_n(t)$ of the 3-phenotype system, with the Ising formulation. (Third row) Dynamical trajectories $x_n(t)$ and $y_n(t)$ of the 3-phenotype system, with the \textit{uncoupled} Ising formulation. (Bottom) Numerical estimation of the progenitor location using the full Ising sysem (black), compared to the analytical approximation (red) derived in \S\ref{section: Supp Timescale}. Phenotype geometry is fixed with $d_x = 2.0$ and $d_y=3.6$.}
    \label{fig:SFig-progenitor-2}
\end{figure}

\begin{figure}[h]
    \centering
    \includegraphics[width=0.85\linewidth]{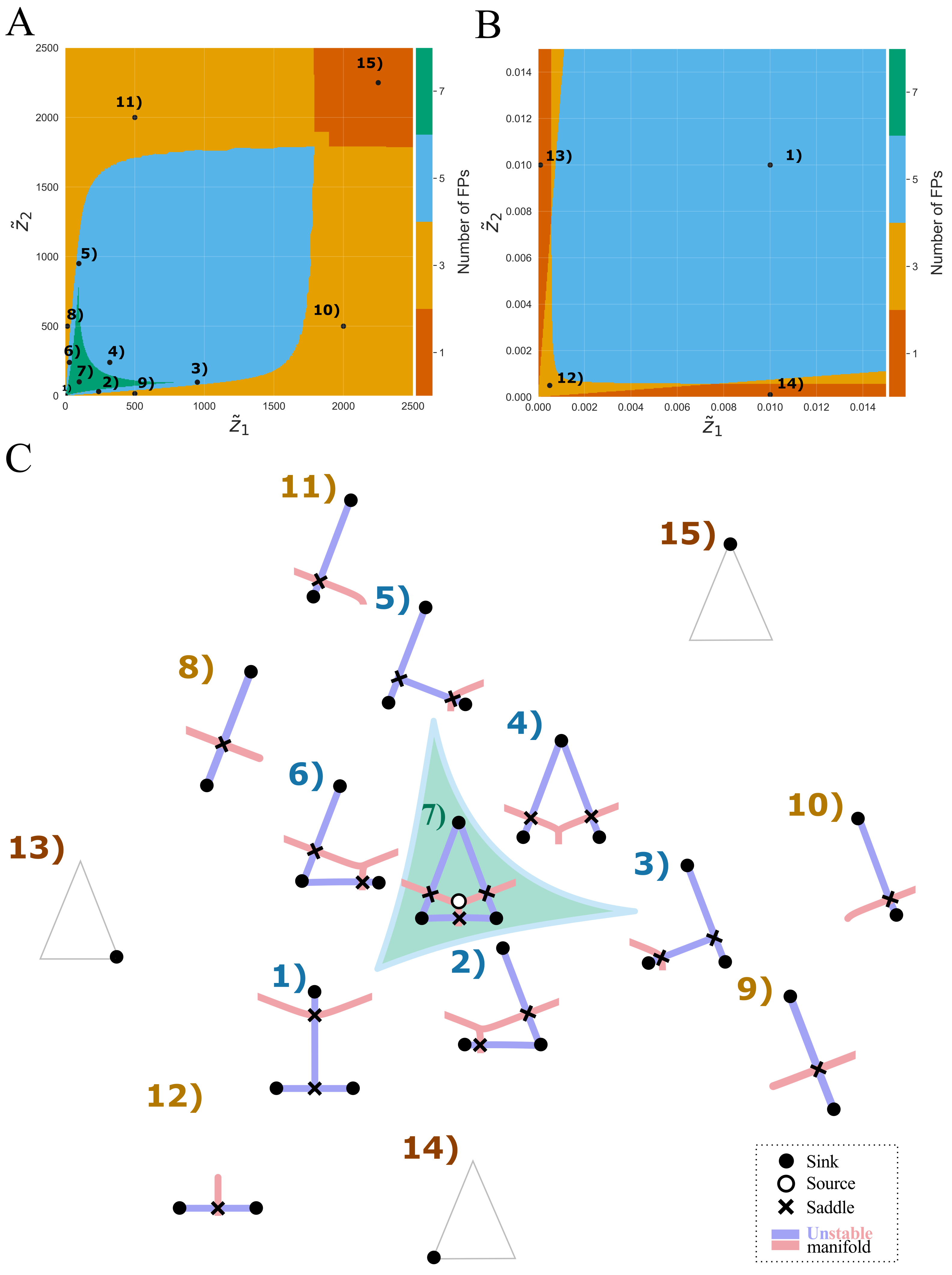}
    \caption[The expanded three-phenotype isosceles catastrophe space.]{Catastrophe space for the three-phenotype isosceles case, with phenotype geometry ($d_x = 2.0$, $d_y=2.6$) and $\beta = 1.5$ as in \ref{fig:fig3_1}. (A) The larger catastrophe space, for $0 \leq \Tilde{z}_1, \Tilde{z}_2 \leq 2500$ (B) Inset of the catastrophe space in (A) for low values $0 \leq \Tilde{z}_1, \Tilde{z}_2 \leq 0.015$. (C) Dynamical geometries for points of interest, labeled 1-15 in (A) and (B). In green (7), the 7-FP geometry. In blue (1-6), the 5-FP geometries, with even-numbered choice-like and odd-numbered flip-like topologies. In orange (8-12), the 3-FP geometries. In red (13-15), the single FP geometries, where the outline of the isosceles triangle is drawn to highlight which phenotype remains.}
    \label{fig:SFig2}
\end{figure}

\begin{figure}[h]
    \centering
    \includegraphics[width=0.99\linewidth]{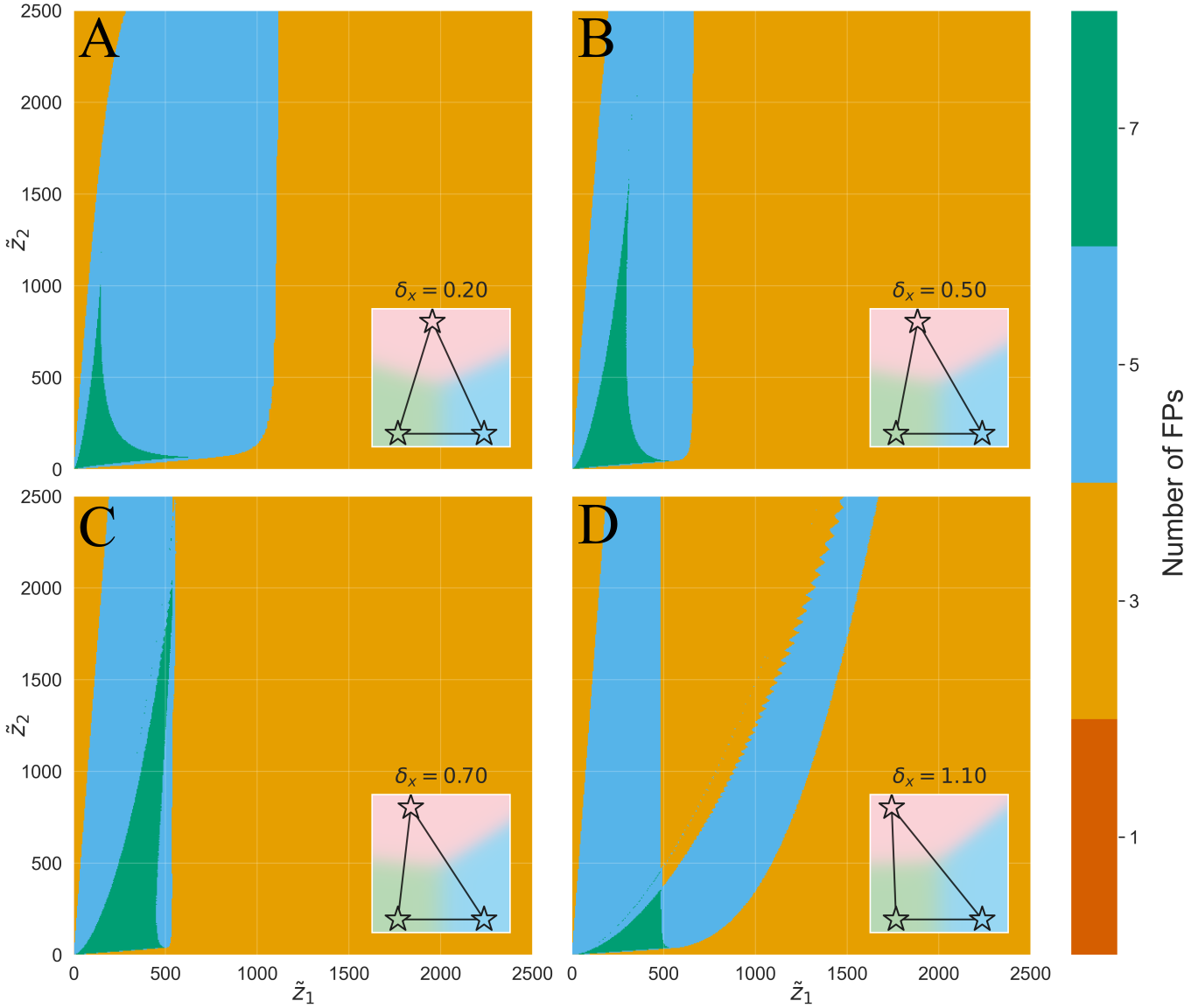}
    \caption[Catastrophe spaces for perturbed three phenotype geometries.]{Catastrophe spaces for perturbed three phenotype geometries. The phenotype geometry in all cases is $\boldsymbol{\xi}_1=(-1.0, -1.3)$, $\boldsymbol{\xi}_2=(+1.0, -1.3)$ and $\boldsymbol{\xi}_3=(-\delta x, -1.3)$, where $\delta_x$ varies between $0.2$ (A), $0.5$ (B), $0.7$ (C) and $1.1$ (D). The phenotype geometry is illustrated in the bottom right corner of each panel. Notice the catastrophe space remains qualitatively unchanged for $\delta_x = 0.2, 0.5$, and the most drastic qualitative change (a novel 3-FP region) appears only for $\delta_x = 1.1$. }
    \label{fig:SFig3}
\end{figure}

\begin{figure}[h]
    \centering
    \includegraphics[width=0.99\linewidth]{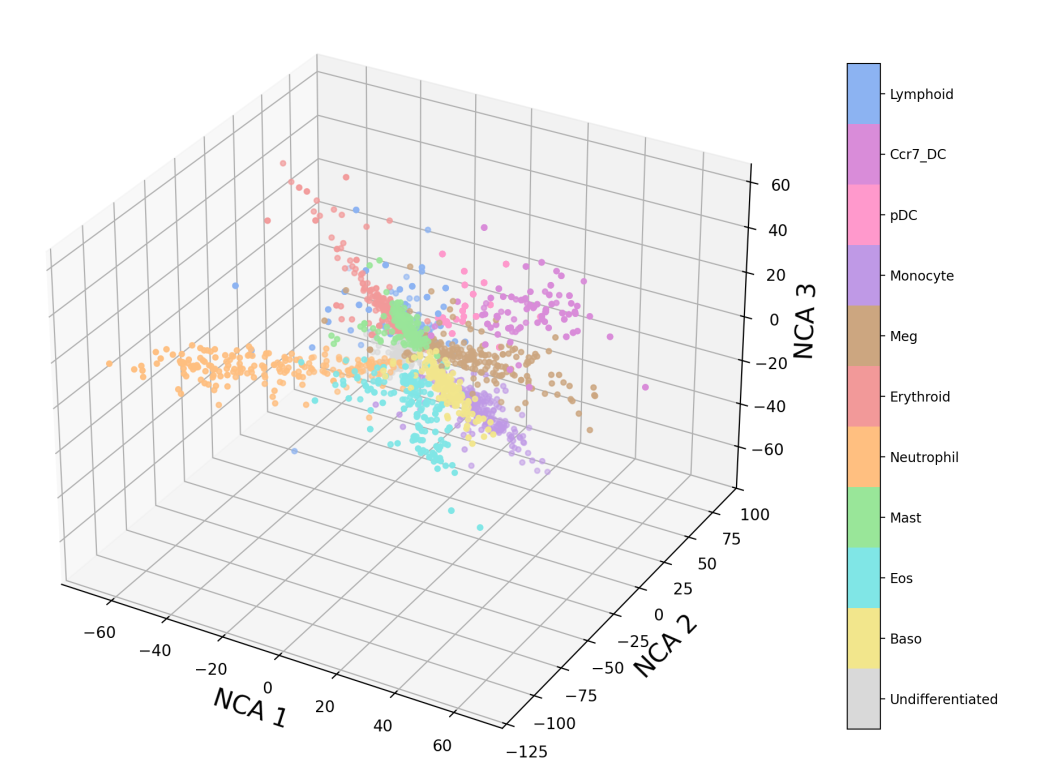}
    \caption[Balanced subset of the Weinreb dataset, projected onto NCA-space.]{Illustration of the Weinreb subset used to generate Fig.~\ref{fig:fig4}. Each celltype has an equal number of profiled cells, namely 200, and are only labeled for visualization purposes. Progenitor/HSCs, colored in grey, are purposefully excluded from the $\boldsymbol{\xi}$'s supplied to the model, but are averaged over, and used as initial conditions for the unit influx.  }
    \label{fig:SFig4}
\end{figure}

\begin{figure}[h]
    \centering
    \includegraphics[width=0.7\linewidth]{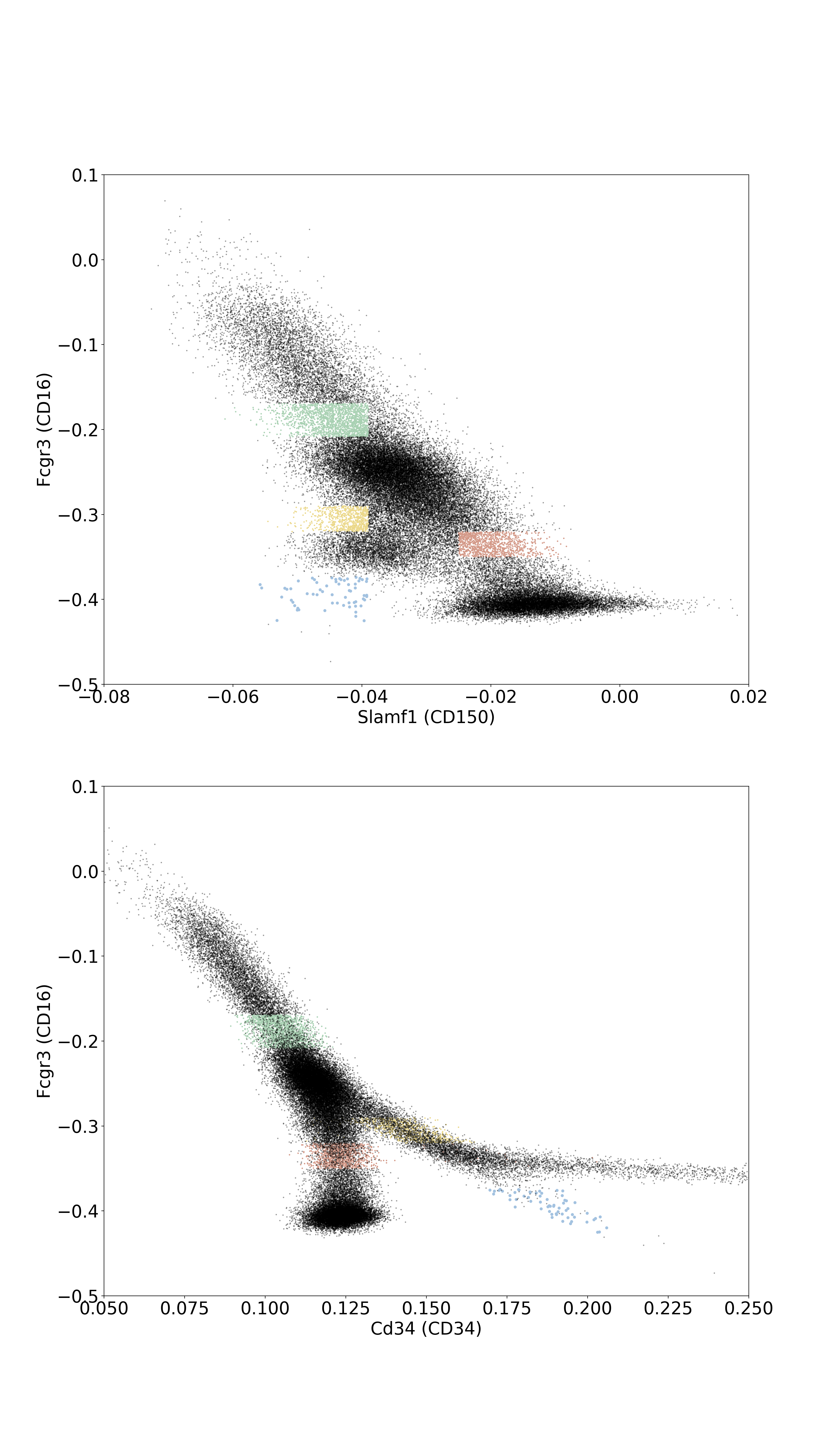}
    \caption[Manual clustering criteria for CMP, CLP, GMP, MEP.]{(A) \textit{Slamf1} and \textit{Fcgr3} gene expression for pre-filtered $\textit{Lin}^{-}\textit{Kit}^{+}$ units.    
    Units are colored with a manual clustering criteria \textit{Slamf1} and \textit{Fcgr3} expression (analagous to CD150 and CD16): CMP (yellow), CLP (blue), GMP (green), MEP (red) and others in black.
    The CLP subcluster is defined as \textit{Slamf1}$^-$\textit{Fcgr3}$^-$, the CMP subcluster is defined as \textit{Slamf1}$^-$\textit{Fcgr3}$^{low}$, the GMP subcluster is defined as \textit{Slamf1}$^-$\textit{Fcgr3}$^+$ and the MEP subcluster is defined as \textit{Slamf1}$^+$\textit{Fcgr3}$^-$. (B) \textit{Cd34} and \textit{Fcgr3} gene expression for pre-filtered $\textit{Lin}^{-}\textit{Kit}^{+}$ units. Interestingly, this is consistent with the experimental CD16 immunophenotype definition of these progenitors, but fails to capture the expected high CD34 expression of GMP. }
    \label{fig:SFig5}
\end{figure}

\begin{figure}[h]
    \centering
    \includegraphics[width=0.99\linewidth]{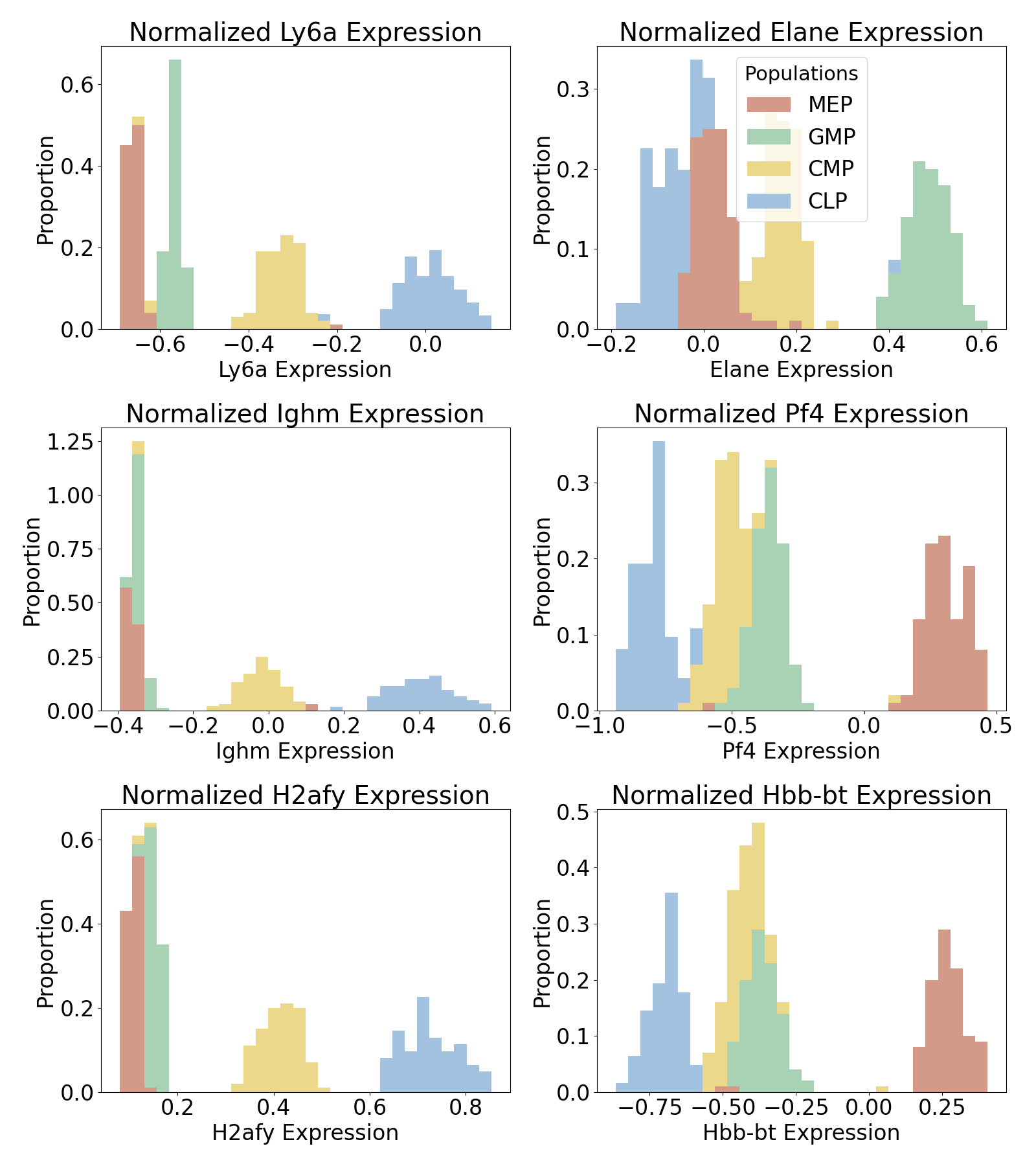}
    \caption[Gene expression histograms for CMP-CLP differential genes.]{Gene expression across identified progenitor subclusters. Normalized distributions of  genetic expression for six genes (\textit{Ly6a}, \textit{Elane}, \textit{Ighm}, \textit{Pf4}, \textit{H2afy} and \textit{Hbb-bt}) across four progenitor populations (CMP, CLP, GMP, MEP) as defined in Fig.~\ref{fig:fig4}. These distributions are directly extracted from the simulation of Fig.~\ref{fig:fig4} and serve as complementary to Fig.~\ref{fig:fig4}G, H.}
    \label{fig:SFig6}
\end{figure}

\begin{figure}[h]
    \centering
    \includegraphics[width=0.99\linewidth]{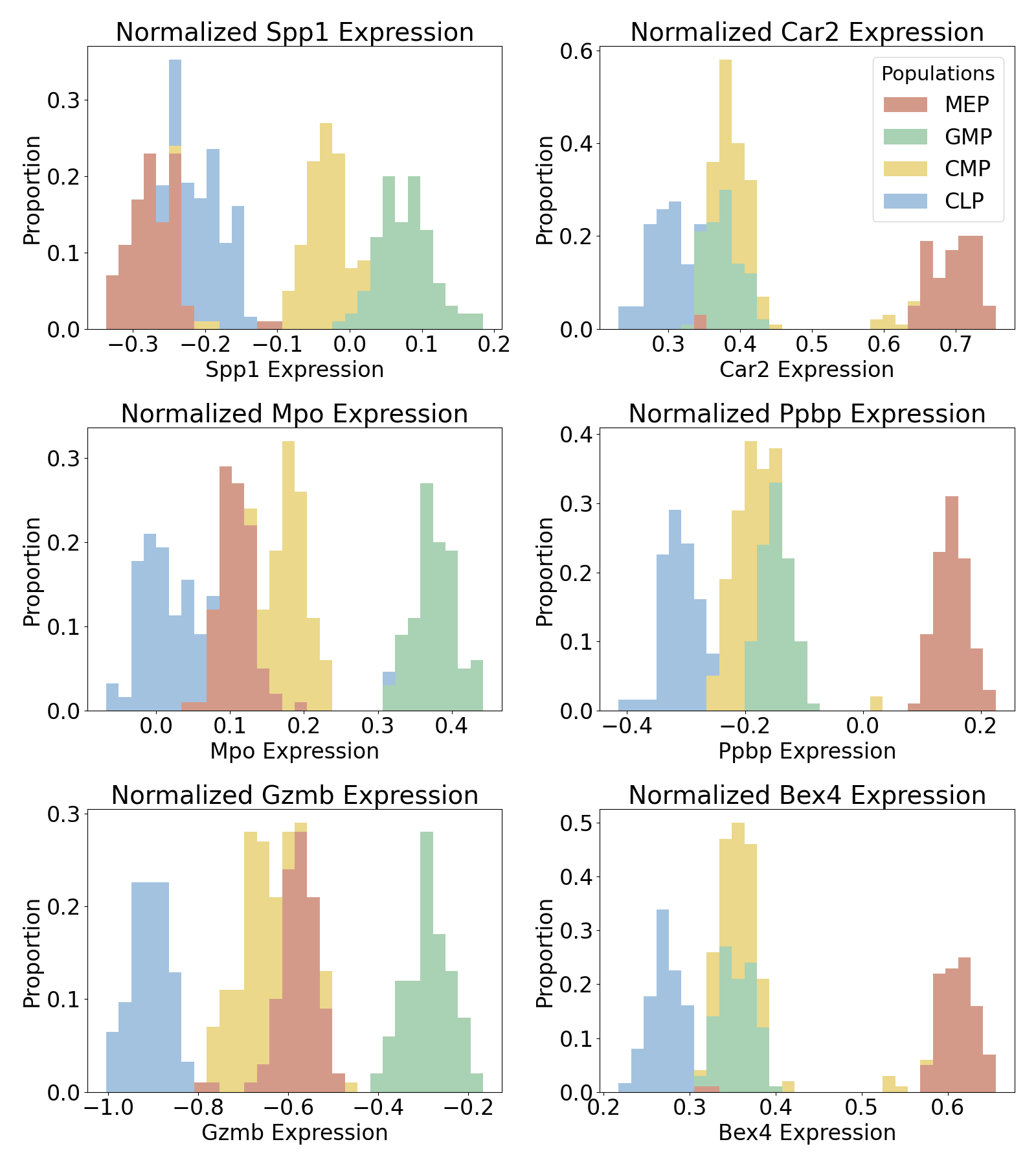}
    \caption[Gene expression histograms for MEP-GMP differential genes.]{Gene expression across identified progenitor subclusters. Normalized distributions of  genetic expression for six genes (\textit{Spp1}, \textit{Car2}, \textit{Mpo}, \textit{Ppbp}, \textit{Gzmb} and \textit{Bex4}) across four progenitor populations (CMP, CLP, GMP, MEP) as defined in Fig.~\ref{fig:fig4}. These distributions are directly extracted from the simulation of Fig.~\ref{fig:fig4} and serve as complementary to Fig.~\ref{fig:fig4}G, H.}
    \label{fig:SFig7}
\end{figure}

\clearpage
\setcounter{figure}{0}
\renewcommand{\thefigure}{M\arabic{figure}}
\renewcommand{\figurename}{\textbf{Supplementary Movie}} 

\makeatletter
\renewcommand{\fnum@figure}{\textbf{\figurename~\thefigure}}
\makeatother

\captionof{figure}[Sandscape dynamics during differentiation, for the 3-phenotype case.]{Movie illustrating the sandscape and differentiation dynamics of 500 units across three phenotypes ($\xi_{blue}=(-1, -1.3)$, $\xi_{green}=(1.0, -1.3)$, and $\xi_{red}=(0, 1.3)$). The underlying energy manifold (sandscape) is computed with eq.~\eqref{eq: energy} and varies spontaneously with the motion of units. Energy isolines are plotted below the manifold in a grey-white colormap. All simulation parameters are exactly as in Fig.~\ref{fig:fig1}B, C.}
\label{mov:Movie 1}

\captionof{figure}[Unit dynamics during differentiation, for the MNIST case.]{Movie illustrating the differentiation of 400 units across 600-phenotypes defined by MNIST handwritten digits (60 per class). Initial conditions are set to the mean digit, up to noise. This movie is complementary to Fig.~\ref{fig:fig1}C, D.}
\label{mov:Movie 2}

\captionof{figure}[Sandscape dynamics for the 3-phenotype case, with a unit-removal perturbation.]{Movie illustrating the sandscape and differentiation dynamics for 500 units and three phenotypes ($\xi_{blue}=(-1, -1.3)$, $\xi_{green}=(1.0, -1.3)$, and $\xi_{red}=(0, 1.3)$). The underlying energy manifold (sandscape) is computed with eq.~\eqref{eq: energy} and varies spontaneously with the motion of units. Energy isolines are plotted below the manifold in a grey-white colormap. Initial dynamics are as in Fig.~\ref{fig:fig1}B, however for $t\in [700, 750]$ units near $\xi_1$ and $\xi_2$ are progressively, and stochastically, pruned until only $\xi_3$-like units remain. All simulation parameters are exactly as in Fig.~\ref{fig:fig3_1}C, E and F.}
\label{mov:Movie 3}

\captionof{figure}[Sandscape dynamics for the 5-phenotype case with unit influx.]{Movie illustrating the dynamics of units differentiating across five phenotypes ($\xi_{green}=(-1.75, -1.0)$, $\xi_{blue}=(1.0, -1.5)$, $\xi_{red}=(0, 1.0)$, $\xi_{orange}=(2.0, 1.0)$ and $\xi_{purple}=(0.9, 3.0)$). 
Units are initialized near the top $\xi_{green}$ attractor, and are added at a constant rate of 3 units/s. The simulation ends with 200,000 units, at which point the sandscape has spontaneously self-organized into a double heteroclinic flip.
The underlying energy manifold (sandscape) is computed with eq.~\eqref{eq: energy} and varies spontaneously with the motion of units. Energy isolines are plotted below the manifold in a grey-white colormap.  All simulation parameters are exactly as in Fig.~\ref{fig:fig3_2}D, E.}
\label{mov:Movie 4}

\captionof{figure}[Unit dynamics during differentiation, for the hematopoiesis case.]{Movie illustrating the differentiation dynamics of units across 2000-phenotypes defined by preprocessed scRNA-sequencing data (200 samples per cell type). Undifferentiated-labelled samples are not included within the phenotypes but are averaged over and used as initial conditions for unit influx. Units are added at a rate of 50 units/s, with 500,000 units at the end of the simulation, at which point the dynamical flow of units traces a tree-like structure with a topology consistent to the classical hematopoiesis differentiation hierarchy.
Units are simulated in a high dimensional PCA-space (100 PCs), obtained from 2006 highly variable genes, but are projected onto an NCA-defined 3D-subspace for visualization. For visualization, units are colored according to the cell type of the nearest phenotype at their terminal state, however the simulation is label-agnostic / unsupervised.
This movie is complementary to Fig.~\ref{fig:fig4}A.}
\label{mov:Movie 5}